\documentclass[11pt,cite,epsfig,psfrag]{article}
\usepackage[utf8]{inputenc}
\usepackage{placeins}
\usepackage{geometry}
 \usepackage{graphicx}
 \usepackage{subfig}
  \usepackage{amsmath} 
  \graphicspath{{YM_Graphs_for_Engine_for_q_12345/}}
  \usepackage[usenames, dvipsnames]{color}
\usepackage{wrapfig}
\usepackage{boxedminipage}

\usepackage{fullpage}
\usepackage{amsmath}
\usepackage{amssymb}
\usepackage{graphicx}
\usepackage{mathrsfs}
\usepackage{wrapfig}
\usepackage{boxedminipage}
\usepackage{epsfig}
\usepackage{setspace}
\usepackage{float}
\usepackage{hyperref}
\usepackage{enumerate}
\usepackage{cite}
\usepackage[font=footnotesize,labelfont=rm]{caption}

\usepackage[numbers,sort&compress]{natbib}

\def\be{\begin{equation}}
\def\ee{\end{equation}}
\def\bea{\begin{eqnarray}}
\def\eea{\end{eqnarray}}

\numberwithin{equation}{section}

 \newcommand{\RN}[1]{%
   \textup{\uppercase\expandafter{\romannumeral#1}}%
 }
\begin{document}

\thispagestyle{empty}
%\baselineskip 20pt
%\rightline{IP/BBSR/2005-12}
%\rightline{\tt hep-th/yymmnnn}

\vskip 2cm

\begin{center}
{\Large \bf A Note on Critical Nonlinearly Charged Black Holes}
\end{center}
%A Note on Power Yang-Mills and Maxwell Black Holes at Criticality

\vskip .2cm

\vskip 1.2cm

\centerline 
{\bf  Pavan Kumar Yerra \footnote{pk11@iitbbs.ac.in} and Chandrasekhar Bhamidipati\footnote{chandrasekhar@iitbbs.ac.in}
}

\vskip 7mm 
\begin{center}{ School of Basic Sciences\\ 
Indian Institute of Technology Bhubaneswar \\ Bhubaneswar 751013, India}
\end{center}

\vskip 1.2cm
\vskip 1.2cm
\centerline{\bf Abstract}
\noindent
Within the extended phase space thermodynamics, we study aspects of power Yang-Mills and power Maxwell black holes at criticality, as the corresponding non-linearity power parameters $\gamma$ and $s$ are varied. On comparison, the approach of efficiency of heat engines to Carnot limit in both the systems is shown to coincide when $\gamma =s=1$. For $\gamma=s> (<) 1$, Maxwell (Yang-Mills) system dominates over Yang-Mills (Maxwell). The motion of charged particles in the critical Power Yang-Mills system is then investigated, together with a study of variation of critical mass to charge ratios with the power $\gamma$.

%\begin{quote}
%\noindent
%\end{quote}
\newpage
\setcounter{footnote}{0}
\noindent

\baselineskip 15pt

\section{Introduction}

Charged black holes in AdS~\cite{Chamblin:1999tk,PhysRevD.60.104026} on the verge of a second order phase transition show a scaling symmetry~\cite{Johnson:2017hxu,Johnson:2017asf}. At this critical point, the thermodynamic quantities scale with respect to charge $q$, i.e., Entropy $S\sim q^2$, Pressure $p\sim q^{-2}$, and Temperature $T\sim q^{-1}$. As a consequence, the geometry of the black hole itself depends on a single parameter $q$, which can be tuned and taken to be large. In a new double scaling limit, where the charge is taken to be large while at the same time nearing the horizon, the geometry of the black hole turns out to be a fully decoupled Rindler space-time, much like the $AdS_2 \times S^2$ space-time obtained from the near horizon limit of extremal black holes. These results might have profound implications for holographic constructions where Rindler space appears in the decoupling limit. The physics of the geometry near the critical point itself is quite interesting from the gravity side. The study of particle dynamics in the effective potential created by the critical black hole is being investigated actively~\cite{Johnson:2017asf,Bhamidipati:2017nau,Hendi:2018wib,Bhamidipati:2018yqy}.\\

\noindent
Black holes at criticality are being actively studied due to the existence of an extended thermodynamic phase space description of charged black holes in AdS~\cite{Chamblin:1999tk,Kubiznak:2012wp,Gunasekaran:2012dq,Cai:2013qga,Kubiznak:2016qmn}, where, there exists a line of first order phase
transitions ending in a second order transition point. In addition to focus on Hawking-Page transition (connecting black holes in AdS to large N gauge theories at finite temperature), Van der Waals transition has attracted considerable attention recently. Holographically~\cite{Johnson:2014yja}, Van der Waals transition might be a journey in the space of field theories (labeled by N, the number of colors in the gauge theory). Thus, varying the cosmological constant in the bulk may be understood as perturbing the dual CFT, triggering a renormalization group flow. This flow is captured in the bulk by Holographic heat engines with black holes as working substances~\cite{Johnson:2014yja}. Novel aspects of this correspondence are being actively explored both from the gravity point of view as well as for applications to the dual gauge theory side~\cite{Dolan:2011xt,Dolan:2010ha,Kastor:2009wy,Caldarelli:1999xj,Sinamuli:2017rhp,Karch:2015rpa,Kubiznak:2014zwa,Johnson:2014yja,Johnson:2015ekr,Johnson:2015fva,Belhaj:2015hha,Bhamidipati:2016gel,Caceres:2015vsa,Chakraborty:2016ssb,Hennigar:2017apu,Johnson:2017ood,Setare2015,Caceres2015,Mo2017,Liu:2017baz,Wei:2016hkm,Sadeghi:2015ksa,Zhang:2016wek,Kubiznak:2016qmn,Sadeghi:2016xal,Zhang:2018vqs,Mo:2017nes,Hendi:2017bys,Wei:2017vqs,Chakraborty:2017weq,Rosso:2018acz,Mo:2018hav,Jafarzade:2018uyz,Johnson:2018amj}. Moreover, at the critical point holographic heat engines have the special property that their efficiency can be very close to that of Carnot engine at finite power\footnote{ See~\cite{PhysRevLett.106.230602,PhysRevLett.111.050601,PhysRevLett.115.090601,PhysRevX.5.031019,PhysRevLett.114.050601,power_of_a_critical_heat,Koning2016,PhysRevLett.117.190601,eff_vs_speed,PhysRevE.96.030102}, for recent discussions on approaching Carnot limit at finite power in thermodynamics and statistical mechanics literature.}, as the scaling parameter $q \rightarrow \infty$~\cite{Johnson:2017hxu}. \\

\noindent
As a generalization of Einstein-Maxwell black holes in AdS, it is interesting to explore new nonlinearly charged systems. Due to infinite self-energy of point like charges in Maxwell's theory~\cite{Born:1934ji,Born:1934gh,1126-6708-2007-12-068,Anninos:2008sj,PhysRevD.78.126007,Seiberg:1999vs,FRADKIN1985123,Metsaev1987207,BERGSHOEFF198770,TSEYTLIN1986391,GROSS198741}, Born and Infeld proposed a generalization when the field is strong, bringing in non-linearities~\cite{Born:1934ji,Born:1934gh, dirac2013lectures, PhysRevD.2.2341}.  In four and higher dimensions, Born-Infeld AdS black holes in extended phase space have been studied\cite{Gunasekaran:2012dq,Zou:2013owa}, with novel results on reentrant phase transition observed in rotating AdS holes ~\cite{Altamirano:2013ane, Altamirano:2013uqa}. Holographic heat engines in Born-Infeld black holes have been constructed in~\cite{Johnson:2015fva,Johnson:2016pfa}, with generalization to Born-Infeld Dilaton case done in~\cite{Bhamidipati:2016gel}.  A novel nonlinear model for electrodynamics is called the power
Maxwell invariant (PMI) theory. In this approach the Lagrangian
density is given by $(-F)^{s},$ being $s$ is an arbitrary rational
number, with a number of novel features and applications in a variety of contexts~\cite{MosqueraCuesta:2004em,MosqueraCuesta:2003dh,Corda:2009xd,AyonBeato:1998ub,AyonBeato:1999rg,AyonBeato:1999ec,AyonBeato:2000zs,AyonBeato:2004ih,PhysRevD.65.063501,Corda:2010ni}. Furthermore, an interesting nonlinear generalization of charged black holes involves a Yang-Mills field coupled to Einstein gravity, where several features in extended thermodynamics have recently been studied~\cite{Zhang:2014eap,Mazharimousavi:2009mb,HabibMazharimousavi:2008ry,ElMoumni:2016eqh,ElMoumni:2018fml}. Unlike Maxwell's case whose range extends to infinity, considerations of Yang-Mills field  inside the nuclei and highly dense matter systems are important justifying their inclusion in applications to Black Holes. In fact, consequences for  thermodynamics of black holes when both the fields are present have actively been pursued~\cite{volkov1989non,Volkov:1989fi,brihaye2007einstein,mazharimousavi2008einstein,bostani2010topological,gao2003first,Devecioglu:2014iia,bellucci2012thermodynamic,HabibMazharimousavi:2008ry,mazharimousavi2008higher,Mazharimousavi:2009mb,PhysRevD.75.027502,hendi2009ricci,hendi2009topological,0264-9381-25-19-195023,PhysRevD.79.044012,Hendi:2010zza,hendi2010slowly,Hendi:2010zz,PhysRevD.82.064040}.
 On the holographically dual side, effects of the nonlinear sources on the
strongly coupled dual gauge theory in the
context of AdS/CFT correspondence have also been reported~\cite{Jing2011,Roychowdhury:2012vj}.\\

\noindent
Motivated by the above developments, in this paper, we study various non-linear effects for black holes in the Einstein-Power Maxwell (PMI) and Einstein Power-Yang-Mills (EPYM) gravity.
It is known that the critical behavior and  Van der Waals like phase transitions exist, depending on charges $q$ and $\bar q$ of the Yang-Mills and Maxwell fields, in addition to the corresponding non-linearity parameters $\gamma$ and $s$, respectively~\cite{Zhang:2014eap,Hendi:2012um}. The plan of the rest of the paper is as follows. Sections-(2.1), (2.2) contain the computation of heat engine efficiency for critical Power Yang-Mills and Power Maxwell black holes, together with a comparison of the two cases. Section-(3) contains results on motion of test particles in the background of critical Power Yang-Mills black holes, where we also present the behavior of critical quantities as the corresponding non-linearity parameter is varied.

 %%%%%%%%%%%%%
\section{Critical Heat Engines for Nonlinear Black Holes} 
%%%%%%%%%%%%%%%%%%%%%
\noindent
We start by defining how the efficiency of holographic heat engines is computed, before proceeding to specific cases.
 Following the analysis of PV criticality in Power Yang-Mills and Power Maxwell Black holes~\cite{Zhang:2014eap,Hendi:2012um}, in this note, we study holographic heat engines in these systems at criticality. Holographic heat engine can be defined for extracting mechanical work from heat energy via the $pdV$ term present in the First  Law of extended black hole thermodynamics~\cite{Johnson:2014yja}, where, the working substance is a black hole solution of the gravity system. One starts by defining a cycle in state space where there is a net input heat flow $Q_H$, a net output heat flow $Q_C$, and a net output work W, such that $Q_H = W + Q_C$. The efficiency of such heat engines can be written in the usual way as $\eta=W/Q_H=1-Q_C/Q_H$. Formal computation of efficiency proceeds via the evaluation of $\int C_p dT$ along those isobars, where~$C_p$ is the specific heat at constant pressure or through an exact formula by evaluating the mass at all four corners as~\cite{Johnson:2015ekr,Johnson:2015fva,Johnson:2016pfa}: 
\begin{equation}
	\eta = 1- \frac{M_3 - M_4}{M_2 - M_1} \, . \label{eq:efficiency-prototype} 
\end{equation}
The equation of state and the mass formula are provided by the black hole in question, which are discussed in the following two subsections.
 %%%%%%%%%%%%%
\subsection{Power Yang-Mills Black Holes:} 
\begin{figure}[h!]
	% \begin{wrapfigure}{l}{0.3\textwidth}
	\begin{center}
		\centering
		\includegraphics[width=3.2in]{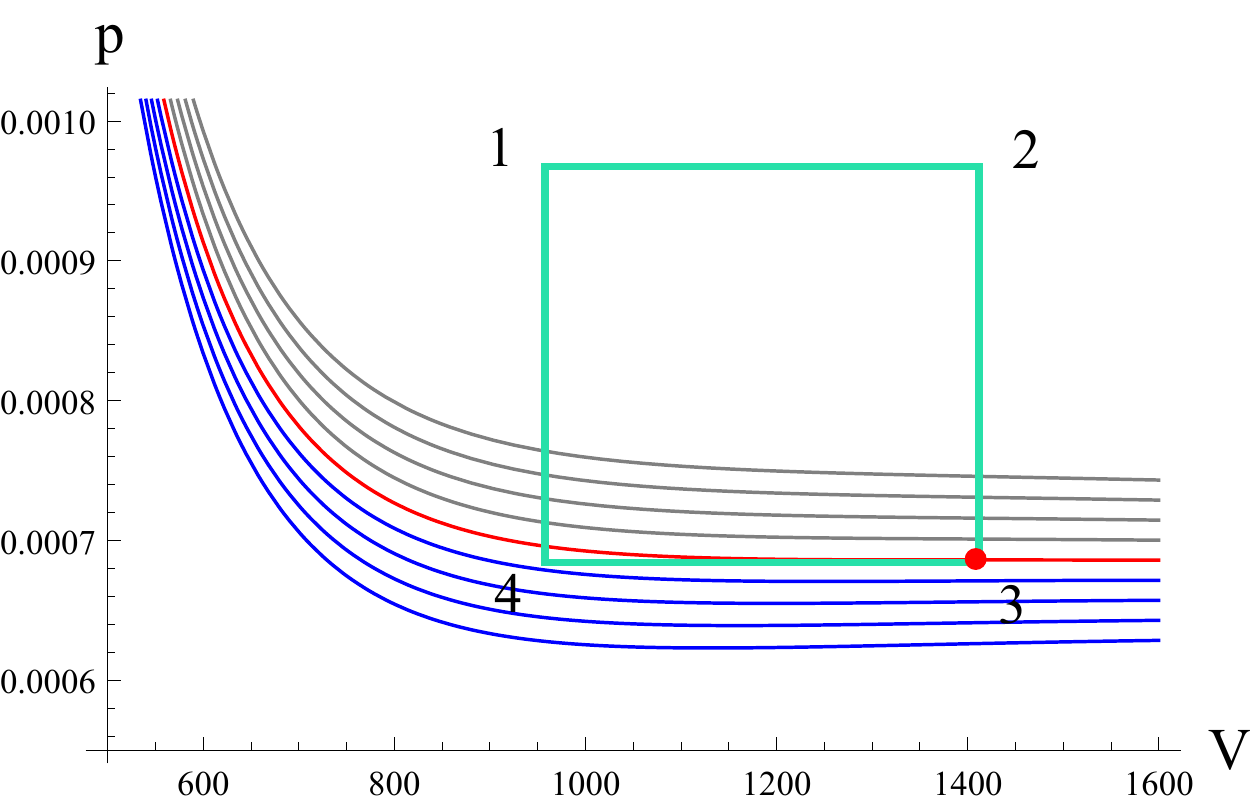}  
		
		\caption{ Sample isotherms for $\gamma = 3$ and $q=10$. The central (red) isotherm is for critical temperature $T_{cr}$, the gray colored isotherms are for $T > T_{cr}$ where the black holes are in unique phase and the blue colored isotherms are for $T < T_{cr}$ where the first order phase transitions are exist. The critical point is highlighted with red dot where  the corner 3 of the engine cycle is  placed.}   \label{fig:isotherms}
		
	\end{center}
	%\end{wrapfigure}
\end{figure}

%%%%%%%%%%%%%%%%%%%%%
Here, we employ  the action for 4-dimensional Einstein--power-Yang--Mills (EPYM) gravity with a cosmological constant $\Lambda$, given by~\cite{Zhang:2014eap,Mazharimousavi:2009mb,HabibMazharimousavi:2008ry,ElMoumni:2018fml}:
\begin{equation}
I=\frac{1}{2}\int\! d^4x \sqrt{-g}\left\{R-2\Lambda -[\textbf{Tr}(F^{(a)}_{\mu\nu}F^{(a)\mu\nu})]^\gamma\right\}\ ,
\label{eq:YM-action}
\end{equation}
where $\textbf{Tr}(.)=\sum_{a=1}^{3}(.)$, $R$ is the Ricci scalar, $\gamma$ is a positive real
parameter and YM (Yang-Mills) field is defined  as
\begin{eqnarray}
&&F_{\mu\nu}^{(a)}=\partial_{\mu}A_{\nu}^{(a)}-\partial_{\nu}A_{\mu}^{(a)}+\frac{1}{2\zeta}
C_{(b)(c)}^{(a)}A_{\mu}^{b}A_{\nu}^{c} \ . 
\end{eqnarray}
Here $C_{(b)(c)}^{(a)}$ represents the structure constants of $3$ parameter
Lie group $G$ and $\zeta$ is a coupling constant, $A_{\mu}^{(a)}$ are the $SO(3)$ gauge
group YM potentials.
The  action admits  the 4-dimensional  EPYM  black hole solution  with negative cosmological constant $\Lambda$ by adopting the metric:
\begin{equation}
ds^2 = -Y( r)dt^2
+ \frac{dr^2}{Y(r)} + r^2 d\Omega^2_2 \, ,
%(d\theta^2+\sin^2\theta d\varphi^2)\ ,
\label{eq:staticform}
\end{equation}
where
\begin{equation}
Y(r)=1-\frac{2m}{r}-\frac{\Lambda}{3}r^2+\frac{(2q^2)^\gamma}{2(4\gamma-3)r^{(4\gamma-2)}}\ .
\label{eq:metric function}
\end{equation}
\noindent
Here, $d\Omega^2_2$ is the metric on a unit 2-sphere with volume $4\pi$ and $q$ is the Yang-Mills charge. We consider the case where solution is valid for $\gamma \neq 0.75$ and  power Yang-Mills term holds the weak energy condition (WEC) when $\gamma > 0$~\cite{Mazharimousavi:2009mb}. 
\vspace{0.6cm}
\\ 
The black hole horizon is located at  $r_+$, the largest positive real root of $Y(r)$, in terms of which   the Hawking temperature $T$, mass $M$ and entropy $S$ of the solution  are given by~\cite{Zhang:2014eap}:

\begin{eqnarray}
	&&T=\frac{Y'(r_+)}{4\pi}=\frac{1}{4\pi r_+} + 2pr_+ - \frac{(2q^2)^\gamma}{8\pi r_+^{(4\gamma
			-1)}}\ , \label{equ:T}\\
	&&M=m=\frac{1}{6}\bigg(8\pi p r_+^{3} + 3r_+^{(3-4\gamma)}\frac{(2q^2)^\gamma}{(8\gamma-6)} + 3r_+ \bigg),\label{equ:M}\\
	&& S= \pi r_+^2 \label{equ:S} \ .
\end{eqnarray}
\noindent
Here, the pressure $p=-\Lambda/(8\pi)$ from extended thermodynamics~\cite{Zhang:2014eap} and the Yang-Mills potential $\Phi_q$ is given by

\begin{equation}
\Phi_q=\frac{2^{(\gamma-1)}q^{(2\gamma-1)}}{(4\gamma-3)}\gamma r_+^{(3-4\gamma)}.
\end{equation}
\noindent
The first law in the extended thermodynamics, where the mass $M$ is understood as the  enthalpy $H$ of the black hole~\cite{Kastor:2009wy}, takes the form as~\cite{Zhang:2014eap}:
\begin{eqnarray}
dM=TdS+\Phi_qdq+Vdp,
\end{eqnarray}
where  the thermodynamic volume  $V \equiv (\frac{\partial M}{\partial p})_{S, q}$ is given by
\begin{equation}
V=\frac{4\pi}{3} r_+^{3}\ = \text{Geometric Volume} .
\label{eq:volume}
\end{equation} 
Now the enthalpy $H(S,p)$ can be expressed as 
\begin{equation}
H\equiv M=\frac{1}{6}\bigg\{8\pi p \bigg(\frac{S}{\pi}\bigg)^{3/2} + 3\bigg(\frac{S}{\pi}\bigg)^{\frac{(3-4\gamma)}{2}}\frac{(2q^2)^\gamma}{(8\gamma-6)} + 3 \sqrt{\frac{S}{\pi}} \bigg\},
\label{eq:enthalpy} 
\end{equation}
while the equation of state $p(V, T)$ can be obtained from the expression of temperature $T$, as:
\begin{equation}
p=\Big(\frac{4\pi}{3V}\Big)^{\frac{1}{3}}\bigg\{\frac{T}{2} - \frac{1}{8\pi}\Big(\frac{4\pi}{3V}\Big)^{\frac{1}{3}} + \frac{(2q^2)^\gamma}{16\pi} \Big(\frac{3V}{4\pi}\Big)^{\frac{(1-4\gamma)}{3}} \bigg\}.
\label{eq:equationofstate}
\end{equation}
\\
This equation of state facilitates the study of phase structure of the black holes in $p-V$ plane.
 As can be seen from the figure (\ref{fig:isotherms}), for a fixed YM charge $q$ and at a given $\gamma$~\cite{Zhang:2014eap},
there exist a first order phase transition line between small and large black holes terminating at the second order critical point.
Above this critical point the black holes exist in a unique phase and hence this phase structure in many aspects resembles the Van der Waals liquid/gas phase transitions. 
\\ \noindent
The stationary point of inflection: $\partial p/\partial V=\partial^2 p /\partial V^2=0$  determines the critical point as~\cite{Zhang:2014eap}:
\begin{eqnarray}
T_{\rm cr}&=& \frac{(2\gamma-1)}{(4\gamma-1)\pi\xi q^\kappa} \ , \nonumber \\
p_{\rm cr} &=& \frac{(2\gamma-1)}{16\pi \gamma \xi^2 q^{2\kappa}} \ , \nonumber \\ 
V_{\rm cr}&=& \frac{4}{3}\pi \xi^3 q^{3\kappa} \ , \nonumber \\
\text{and} \, \, \, \, r_{\rm cr}&=& \xi q^\kappa , \nonumber
\label{eqn:YM critical point}
\end{eqnarray}
\begin{equation}
  \text{where} \, \, \, \quad \kappa = \frac{\gamma}{(2\gamma - 1)} \quad \text{and} \quad \xi = \big[\gamma(4\gamma-1)2^\gamma\big]^{\frac{1}{(4\gamma-2)}}. 
\end{equation}

\noindent
 Further, the specific heats at constant volume and pressure  are given by~\cite{Zhang:2014eap,Dolan:2011xt}:

\begin{equation}
C_V=0 \ ; \, \, \,   C_p = 2S\Big\{\frac{2\pi S^{(2\gamma)}(8pS+1)-S(2\pi^2q^2)^\gamma}{2\pi S^{(2\gamma)}(8pS-1)+S(4\gamma-1)(2\pi^2q^2)^\gamma}\Big\} .
\end{equation}
\noindent
Since $C_V=0$, as a natural choice~\cite{Johnson:2014yja}, we choose the engine cycle as a rectangle in $p-V$ plane, and the equation (\ref{eq:efficiency-prototype}) serves in computing the efficiency. It was stated in~\cite{PhysRevLett.114.050601,power_of_a_critical_heat} that, to approach the Carnot limit with non vanishing power, it is useful to place the engine at the  verge of the second order critical point (or even close to critical point).
 This novel feature was even observed  in the context of charged-AdS black holes~\cite{Johnson:2017hxu} and also in Gauss-Bonnet black holes~\cite{Bhamidipati:2017nau} when the corresponding parameter (charge $q$ or Gauss-Bonnet coupling $\alpha$) is taken to be large.
\vspace{0.3cm} \\ This idea of working with critical point motivates us to place our cycle close to the critical point and also to study the consequences of YM power $\gamma$ on the efficiency of the engine, as compared to the case of standard charged black holes in AdS. Following the prescription in~\cite{Johnson:2017hxu}, we place the critical point at the corner 3 (see fig. \ref{fig:isotherms}) and choose the boundaries of the cycle\footnote{ placing  the critical point at other corners and choosing different boundaries is also permissible, but we opt for this choice in~\cite{Johnson:2017hxu} for later comparison.} such that
\begin{eqnarray}
p_1 &= & p_2  =  3p_{\rm cr}/2,  \nonumber \\
p_3 &= &p_4 =p_{\rm cr}, \nonumber \\ 
V_2 &=& V_3=V_{\rm cr},  \nonumber \\ 
\text{and} \, \, \, \,  V_1 &=& V_4= V_{\rm cr}\big(1-\frac{L}{q^\kappa}\big), 
\end{eqnarray}  
\noindent
where  L is a constant with dimensions of $q^\kappa$ such that $ 0 <L<q^\kappa$ (without loss of generality, we set $ L = 1$).
 The  area of the cycle gives the work $W$ done by the engine as:
 \begin{equation}
  W = \Delta p \Delta V =\frac{\xi(2\gamma-1)}{24\gamma}.
 \end{equation} This work is independent of YM charge $q$ and finite, while the heat $Q_H$ entering the cycle can be calculated from the difference in enthalpies as\footnote{It is important to check the validity of physical quatities such as Mass in various ranges of $\gamma$, while computing the efficiency. }:
 \begin{eqnarray}
 Q_H & = & M_2 -M_1  \nonumber \\
 &=& \frac{\xi q^\kappa}{6}\Bigg\{\frac{(6\gamma-3)}{4\gamma q^\kappa} + \frac{3\Big[1-\big(1-\frac{1}{q^\kappa}\big)^{(3-4\gamma)/3}\Big]}{\gamma(4\gamma-1)(8\gamma-6)} + 3\Big[1-\Big(1-\frac{1}{q^\kappa}\Big)^{\frac{1}{3} }\Big]\Bigg\}.
 \label{eq: Exact	QH } \end{eqnarray} 
\noindent
Its large $q$ expansion is:
\begin{equation}
Q_H= \frac{(1-22\gamma+40\gamma^2)\xi}{24\gamma (4\gamma-1)} + \frac{(2\gamma-1)\xi}{9(4\gamma-1)}\Big(\frac{1}{q^\kappa}\Big) + \frac{4(2\gamma-1)\xi}{81(4\gamma-1)}\Big(\frac{1}{q^{2\kappa}}\Big) + O\Big( {\frac{1}{q^{3\kappa}}}\Big)\ ,
\end{equation}
where as the  large $q$ expansion of efficiency $\eta=W/Q_H$  is:
\begin{eqnarray}
\eta &=&\Big(\frac{1-4\gamma}{ 1-20\gamma}\Big) - \frac{8\gamma(4\gamma-1)}{3(1-20\gamma)^2}\Big(\frac{1}{q^\kappa}\Big) - \frac{32\gamma(4\gamma-1)(14\gamma-1)}{27(20\gamma-1)^3}\Big(\frac{1}{q^{2\kappa}}\Big) \nonumber \\
 &&  + \, \,  \frac{2\gamma(4\gamma-1)(1600\gamma^3-7408\gamma^2+908\gamma-29)}{81(1-20\gamma)^4}\Big(\frac{1}{q^{3\kappa}}\Big) + O\Big( {\frac{1}{q^{4\kappa}}}\Big)\ .
\end{eqnarray}
To benchmark our engine against the Carnot engine (which sets an upper bound on efficiency irrespective of the nature of working substance) highest and lowest  temperatures ($T_H, T_C$) between which our engine works is supplied by the equation of state as:
\begin{equation}
T_2=T_H = \frac{1}{4\pi\xi q^\kappa}\Big\{1+\frac{3}{4\kappa} - \frac{1}{2\gamma(4\gamma-1)}\Big\}  \, , 
\end{equation}
and  the large $q$ expansion  for $T_C$  is:
\begin{equation} 
T_4 = T_C =  \frac{(1-2\gamma)}{\pi \xi(1-4\gamma)}\Big(\frac{1}{q^\kappa}\Big) + \frac{(1-2\gamma)}{162\pi \xi}\Big(\frac{1}{q^{4\kappa}}\Big) - \frac{(2\gamma-1)(17+4\gamma)}{1944\pi \xi}\Big(\frac{1}{q^{5\kappa}}\Big)  + O\Big( {\frac{1}{q^{6\kappa}}}\Big)\ .
\end{equation}
These temperatures yield the Carnot efficiency $\eta_{\rm C}^{\phantom{C}}$ at large $q$ as:
\begin{equation}
\eta_{\rm C}^{\phantom{C}}=1-\frac{T_C}{T_H} = \Big(\frac{1-4\gamma}{ 1-20\gamma}\Big) + \frac{8\gamma(4\gamma-1)}{81(20\gamma-1)}\Big(\frac{1}{q^{3\kappa}}\Big) + \frac{2\gamma(16\gamma^2+64\gamma-17)}{243(20\gamma-1)}\Big(\frac{1}{q^{4\kappa}}\Big)  +  O\Big( {\frac{1}{q^{5\kappa}}}\Big).
  \end{equation}

  \begin{figure}[h]
  	% \begin{wrapfigure}{l}{0.3\textwidth}
  	\begin{center}
  		{\centering
  			\subfloat[]{\includegraphics[width=2.8in]{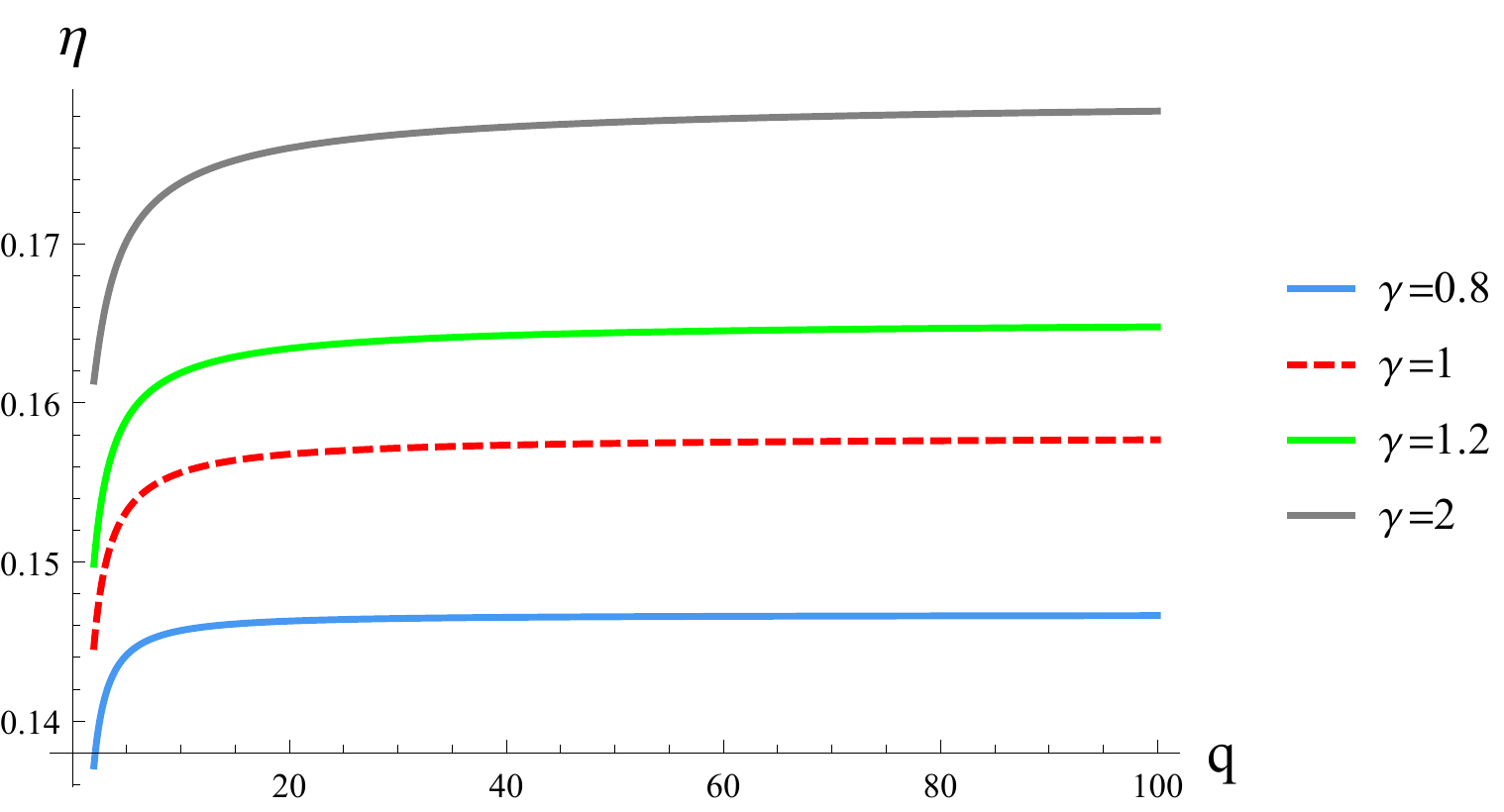} } \hspace {1.3cm}
  			\subfloat[]{\includegraphics[width=2.8in]{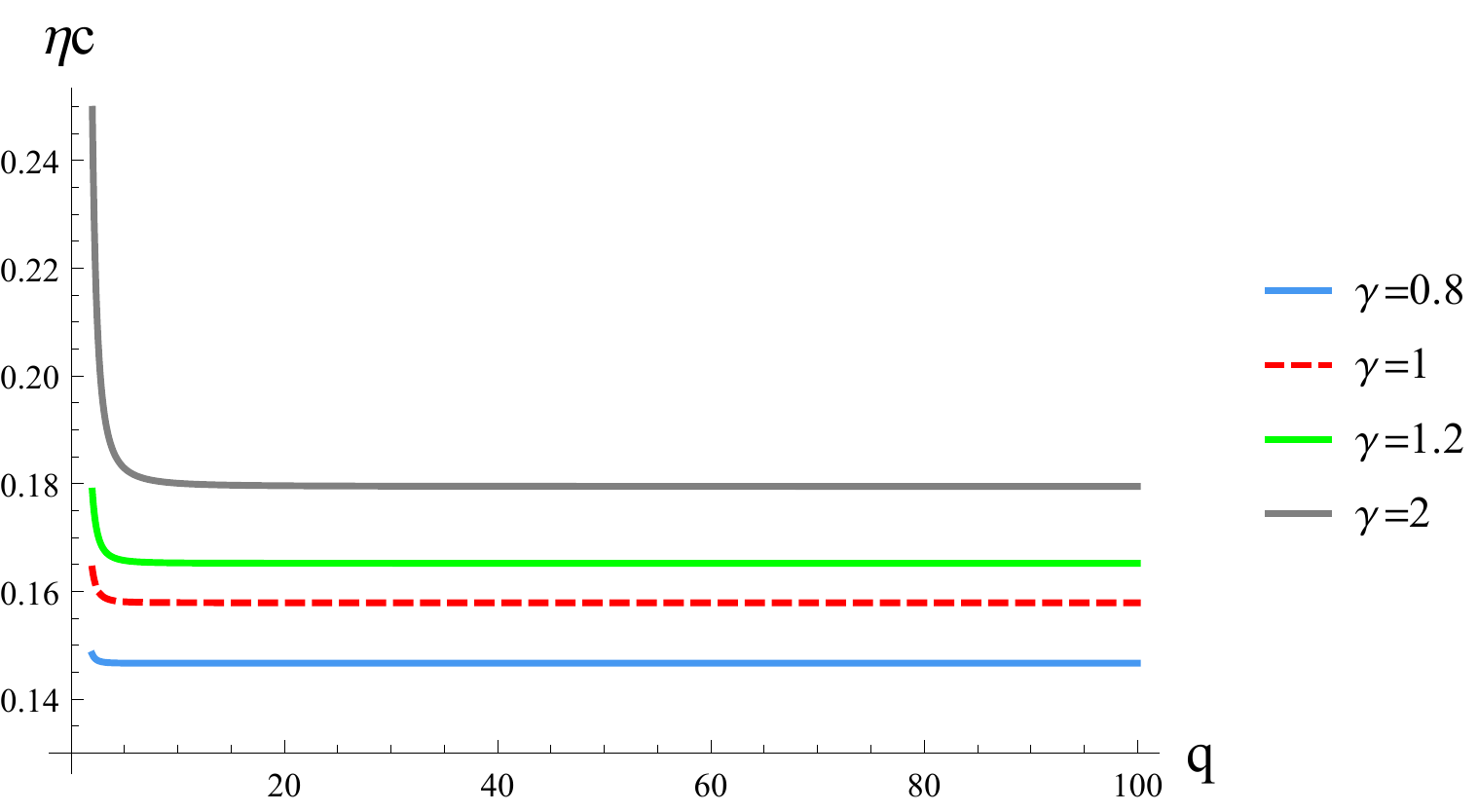} }
  			\caption{ The behaviour of (a) $\eta$ and  (b) $ \eta_{\rm C}$    with YM charge $q$ at different $\gamma$.}   \label{fig:YM plots of eta and etac at gamma 12345}
  		}
  	\end{center}
  	%\end{wrapfigure}
  \end{figure}
  
  \begin{figure}[h]
  	% \begin{wrapfigure}{l}{0.3\textwidth}
  	\begin{center}
  		{\centering
  			\includegraphics[width=2.9in]{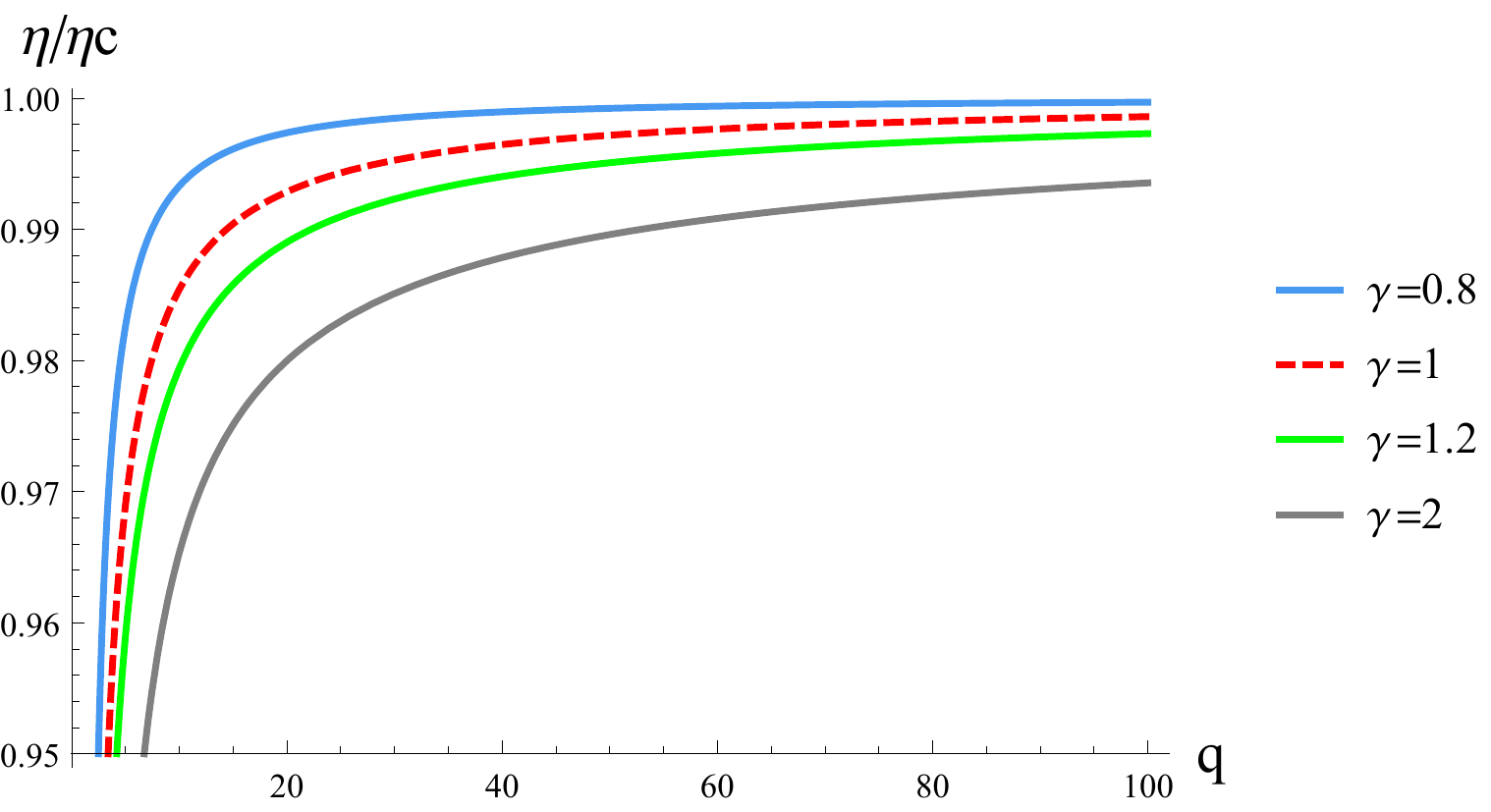}    		
  			\caption{ The behaviour of   the ratio $\eta/\eta_{\rm C}$  with YM charge $q$ at different $\gamma$.}   \label{fig:YM plot of ita by itac for gamma 12345}
  		}
  	\end{center}
  	%\end{wrapfigure}
  \end{figure}
\noindent 
Figure (\ref{fig:YM plots of eta and etac at gamma 12345}) shows efficiency of the engine as well as Carnot efficiency as $\gamma$ is varied. Figure (\ref{fig:YM plot of ita by itac for gamma 12345}) illustrates that, for a given $\gamma$, the ratio $\eta/\eta_{\rm C}$ approaches to unity when the YM charge $q$ takes larger values. Moreover, for a given YM charge $q$, the approach of the ratio $\eta/\eta_{\rm C}$ towards unity  is high for lower values  of $\gamma$.\\

\noindent
The dependence of pressures in the engine  on the critical pressure $p_{cr}$, helps us to estimate the time $\widehat{\tau}$ taken to complete a cycle~\cite{Johnson:2017hxu}. As  $p_{cr} \sim q^{-2\kappa}$,  the estimation of the time turns out to be  \, $\widehat{\tau} \sim q^{2\kappa}$.
 Finite time $\widehat{\tau}$, for a finite $q$, leads to finite power  $ (= W/\widehat{\tau})$ (as the work $W$ is finite).
However, when the YM charge $q$ go towards infinite where $ \eta=\eta_{\rm C}$,  power  $ (= W/\widehat{\tau})$ vanishes in this limit. This even satisfies the \textit{universal trade off relation} between power and efficiency, given by~\cite{PhysRevLett.117.190601,eff_vs_speed}:
  \begin{equation}
  \frac{W}{\widehat{\tau} }\leq{\bar\Theta} \frac{\eta(\eta_{\rm C}^{\phantom{C}}-\eta)}{T_C}\ , \label{trade off relation}
  \end{equation}
  where ${\bar\Theta}$ is a  constant  associated with the engine.  
  The large $q$ expansion of the right hand side quantity  in eqn.(\ref{trade off relation}) (on dividing by ${\bar\Theta}$) is:
  \begin{equation} 
  	\frac{\eta(\eta_{\rm C}-\eta)}{T_C}  =  \frac{8\pi \gamma(4\gamma-1)^3\xi}{3(2\gamma-1)(20\gamma-1)^3} + \frac{32\pi \gamma(4\gamma-1)^3(8\gamma-1)\xi}{27(1-20\gamma)^4(2\gamma-1)}\Big(\frac{1}{q^{\kappa}}\Big)  + O\Big( {\frac{1}{q^{2\kappa}}}\Big)\ .
  	\end{equation}
 	
%\subsection*{}
%\underline{\textbf{At $\gamma = 1$:}}
%\vspace{0.3cm}
%\\
\noindent
 When we set the Yang-Mills power $\gamma$ to be unity, the above results are exactly same as those of the critical heat engine for 4-dimensional Reissner-Nordstrom (RN)  black hole (with $L=1$, there)~\cite{Johnson:2017hxu}, where the Yang-Mills charge plays the role of the Maxwell charge (see Appendix for details).
 %%%%%%%%%%%%%
\subsection{Power Maxwell Black Holes:} 
%%%%%%%%%%%%%%%%%%%%%
Now we would like to compare our results with those of Power Maxwell charged black holes. 
The corresponding action and the metric are given respectively, as~\cite{Hendi:2012um}:
\begin{equation}
I= -\frac{1}{16\pi}\int\! d^4x \sqrt{-g}\Big\{R-2\Lambda + (-F_{\mu\nu}F^{\mu\nu})^s \Big\}\ ,
\end{equation}
\begin{equation}
ds^2 = -f( r)dt^2
+ {dr^2\over f(r)} + r^2 d\Omega^2_2 \ ,
\end{equation}
where the metric function $f(r)$, gauge potential one form $A$ and electromagnetic field two form $F$ are given by:
\begin{eqnarray}
f(r) & = & 1 -  \frac{r^2\Lambda}{3} -\frac{m}{r} + \frac{a(2s-1)^2\bar{q}^{2s}}{2(3-2s) r^{2/(2s-1)}} \ ,  \\
A &=& - \bar{q} r^{(2s-3)/(2s-1)}dt \ , \\
F &=& dA. \\
\text{where}, \, \, \, \,  a & = & \Big[\frac{2(2s-3)^2}{(2s-1)^2}\Big]^s.
\end{eqnarray}
 The power $s \neq \frac{3}{2}$ is restricted to $s> \frac{1}{2}$~\cite{Hendi:2012um}, and the parameters $m$ and $\bar{q}$ are, respectively, related to mass $M$ and electric charge $Q$ of the black hole.
 In term of horizon radius $r_+$ of the black hole, where $f(r=r_+)=0$, the thermodynamic quantities of the black hole are given by~\cite{Hendi:2012um}:  
\begin{eqnarray}
\text{Temperature T} & = & \frac{f'(r_+)}{4\pi} = \frac{1}{4\pi r_+}\Big\{1+ 8\pi pr_+^2 - \frac{(2s-1)a\bar{q}^{2s}}{2r_+^{2/(2s-1)}}\Big\} \ ,\\
\text{Mass M} &=& \frac{m}{2} = \frac{r_+}{2} \Big\{1+ \frac{8\pi pr_+^2}{3} + \frac{a(2s-1)^2\bar{q}^{2s}}{(6-4s)r_+^{2/(2s-1)}}\Big\} \ ,\\
\text{charge Q} &=& 2^{(s-1)}(2s-1)s\bigg[\frac{(3-2s)\bar{q}}{2s-1}\bigg]^{(2s-1)} \ ,\\
\text{electric potential $ \Phi $} &=& \frac{\bar{q}}{r_+^{(3-2s)/(2s-1)}} \ , \\
\text{entropy S} &=& \pi r_+^2 \ , \, \, \text{and} \\
\text{Thermodynamic Volume V} &=& \frac{4\pi}{3} r_+^3.
\end{eqnarray}
Now enthalpy $M(S,p)$ and equation of state $p(V,T)$ 
%in canonical ensemble (fixed $Q$) 
are given by:
\begin{eqnarray}
M (S, p) &=& \frac{1}{2}\sqrt{\frac{S}{\pi}}\bigg\{1+\frac{8pS}{3} + \frac{a(2s-1)^2\bar{q}^{2s}}{(6-4s)\big( \frac{S}{\pi}\big)^{1/(2s-1)}} \bigg\} \ , \\
p(V,T) &=& \Big(\frac{4\pi}{3V}\Big)^{\frac{1}{3}} \bigg\{ \frac{T}{2} -\frac{1}{8\pi}\Big(\frac{4\pi}{3V}\Big)^{\frac{1}{3}} + \frac{(2s-1)a\bar{q}^{2s}}{16\pi}\Big(\frac{4\pi}{3V}\Big)^{(\frac{2s+1}{6s-3})} \bigg\}.
\end{eqnarray}
This equation of state exhibits Van der Waals behaviour (similar to Yang-Mills case) with the critical point~\cite{Hendi:2012um}:
\begin{equation}
\quad p_{\rm cr}=\frac{1}{16\pi s \chi^2 \bar{q}^{2b}}\ , \quad V_{\rm cr}= \frac{4}{3}\pi \chi^3 \bar{q}^{3b} \ ,   \quad T_{\rm cr}= \frac{1}{\pi (2s+1)\chi\bar{q}^b}\ , 
\end{equation}
where \, \,  $ r_{\rm cr}= \chi \bar{q}^b $,  \, \, $\chi = \bigg[\frac{as(2s+1)}{(2s-1)}\bigg]^{\big(\frac{2s-1}{2}\big)} $ and \, \,  $ b= s(2s-1)$.
One obtains the specific heats as~\cite{Hendi:2012um}:
\begin{equation}
C_V=0 \ ; \, \, \,   C_p= \frac{2S\bigg[ 2S^{\frac{1}{2s-1}}(8pS+1)-a\pi^{\frac{1}{2s-1}}\bar{q}^{2s}(2s-1) \bigg]}{\bigg[ 2S^{\frac{1}{2s-1}}(8pS-1)+a\pi^{\frac{1}{2s-1}}\bar{q}^{2s}(2s+1) \bigg]}  \ .
\end{equation}
The work $W = \frac{\chi}{24s} $ is finte for the rectangular cycle with the following boundaries:
\begin{eqnarray}
p_3 &= &p_4 =p_{\rm cr}, \nonumber \\ 
p_1 &= & p_2  =  3p_{\rm cr}/2,  \nonumber \\
V_2 &=& V_3=V_{\rm cr},  \nonumber \\ \text{and} \, \, \, \,  V_1 &=& V_4= V_{\rm cr}\Big(1-\frac{L}{\bar{q}^b}\Big),  
\end{eqnarray} 
where $L$ is a constant having dimensions of $\bar{q}^b$ such that  $0<\frac{L}{\bar{q}^b}<1$, and we set $L=1$.
Now the heat inflow $Q_H = M_2-M_1$ is:
\begin{equation}
Q_H = \frac{\chi\bar{q}^b}{2}\bigg\{1-\Big(1-\frac{1}{\bar{q}^b}\Big)^{1/3} + \frac{1}{4s\bar{q}^b} + \frac{a(2s-1)^2}{(6-4s)\chi^{\frac{2s}{b}}} \bigg[1-\Big(1-\frac{1}{\bar{q}^b}\Big)^{\frac{b-2s}{3b}}\bigg] \bigg\} \ ,
\end{equation}
and its large $\bar{q}$ expansion is:
\begin{equation}
Q_H= \frac{(1+18s)\chi}{24s(1+2s)} + \frac{\chi}{9(1+2s)} \left( \frac{1}{\bar{q}^{b}} \right) + \frac{4\chi}{81(1+2s)} \left( \frac{1}{\bar{q}^{2b}} \right)+ \frac{(54s-29)\chi}{972(4s^2-1)} \left( \frac{1}{\bar{q}^{3b}} \right) +O \left( \frac{1}{\bar{q}^{4b}} \right)\ ,
\end{equation}
while efficiency $\eta$ at large $\bar{q}$ is:
\begin{equation}
\eta =\Big(\frac{1+2s}{1+18s}\Big) - \frac{8s(1+2s)}{3(1+18s)^2} \left( \frac{1}{\bar{q}^{b}} \right) - \frac{32s(1+2s)(1+12s)}{27(1+18s)^3} \left( \frac{1}{\bar{q}^{2b}} \right) +O \left( \frac{1}{\bar{q}^{3b}} \right).
\end{equation}
The  highest and lowest temperatures ($T_H$ and $T_C$, respectively) of the cycle are:
\begin{eqnarray}
T_H = T_2 &=& \frac{1}{4\pi\chi \bar{q}^b}\Big\{ 1+ \frac{3}{4s} - \frac{ab}{2s\chi^{2s/b}}\Big\} \, , \\
T_C= T_4 &= &\frac{1}{\pi\chi(1+2s)} \left( \frac{1}{\bar{q}^{b}} \right) + \frac{1}{162\pi\chi(1-2s)} \left( \frac{1}{\bar{q}^{4b}} \right) \nonumber \\ 
&& + \frac{(17-38s)}{1944\pi(1-2s)^2\chi} \left( \frac{1}{\bar{q}^{5b}} \right) +O \left( \frac{1}{\bar{q}^{6b}} \right) \, \, \, \, \, \, \, \text{(at large $\bar{q}$)},
\end{eqnarray}
where the Carnot efficiency $ \eta_{\rm C}^{\phantom{C}}$, at large $\bar{q}$, turns out to be:
\begin{equation}
 \eta_{\rm C}^{\phantom{C}} =\Big(\frac{1+2s}{1+18s}\Big) + \frac{8s(1+2s)}{81(2s-1)(1+18s)} \left( \frac{1}{\bar{q}^{3b}} \right) + \frac{2s(76s^2+4s-17)}{243(1-2s)^2(1+18s)} \left( \frac{1}{\bar{q}^{4b}} \right) +O \left( \frac{1}{\bar{q}^{5b}} \right).
\end{equation}

 \begin{figure}[h]
 
 	\begin{center}
 		{\centering
 			\subfloat[]{\includegraphics[width=2.8in]{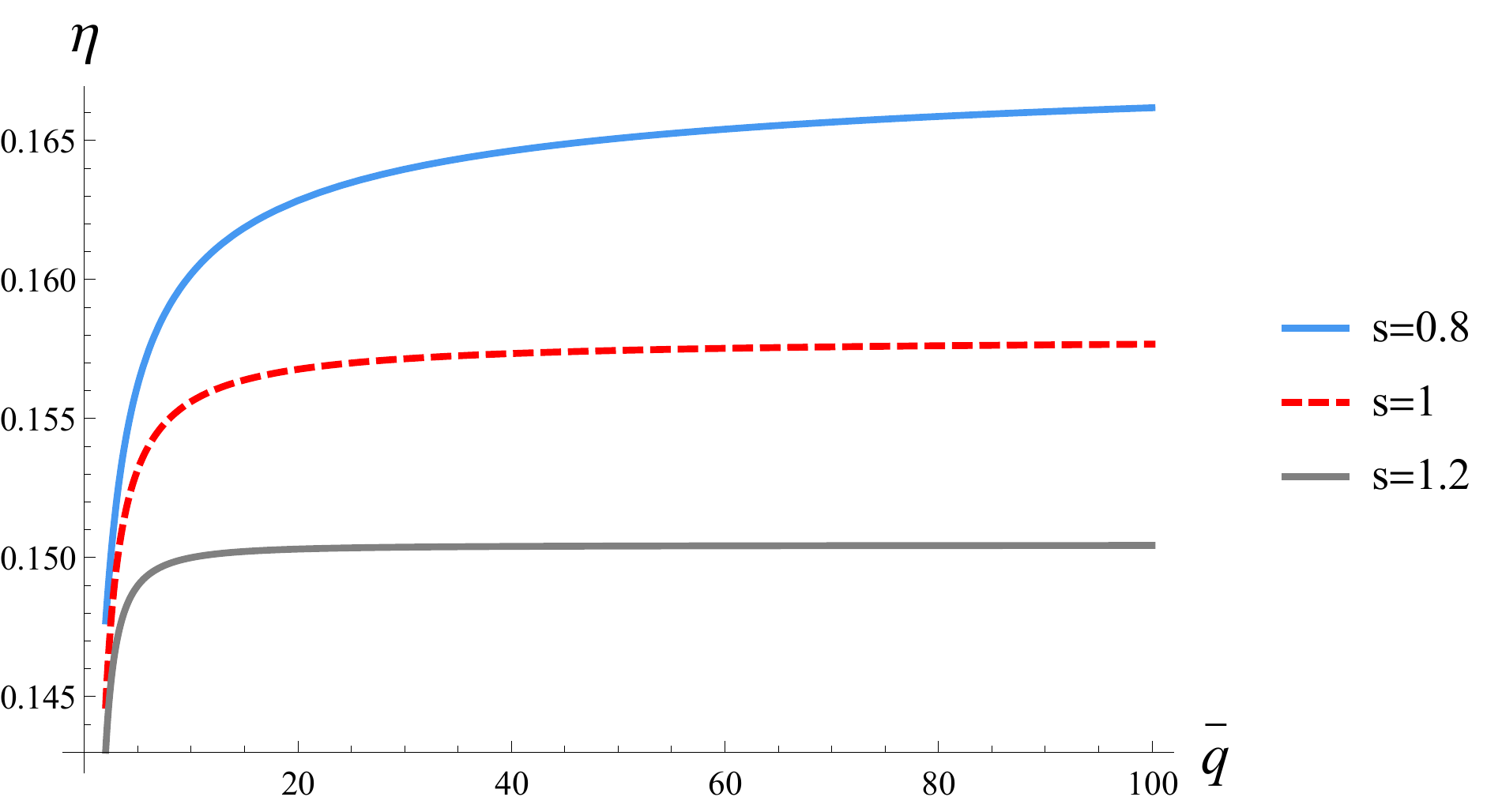} } \hspace {1.0cm}
 			\subfloat[]{\includegraphics[width=2.8in]{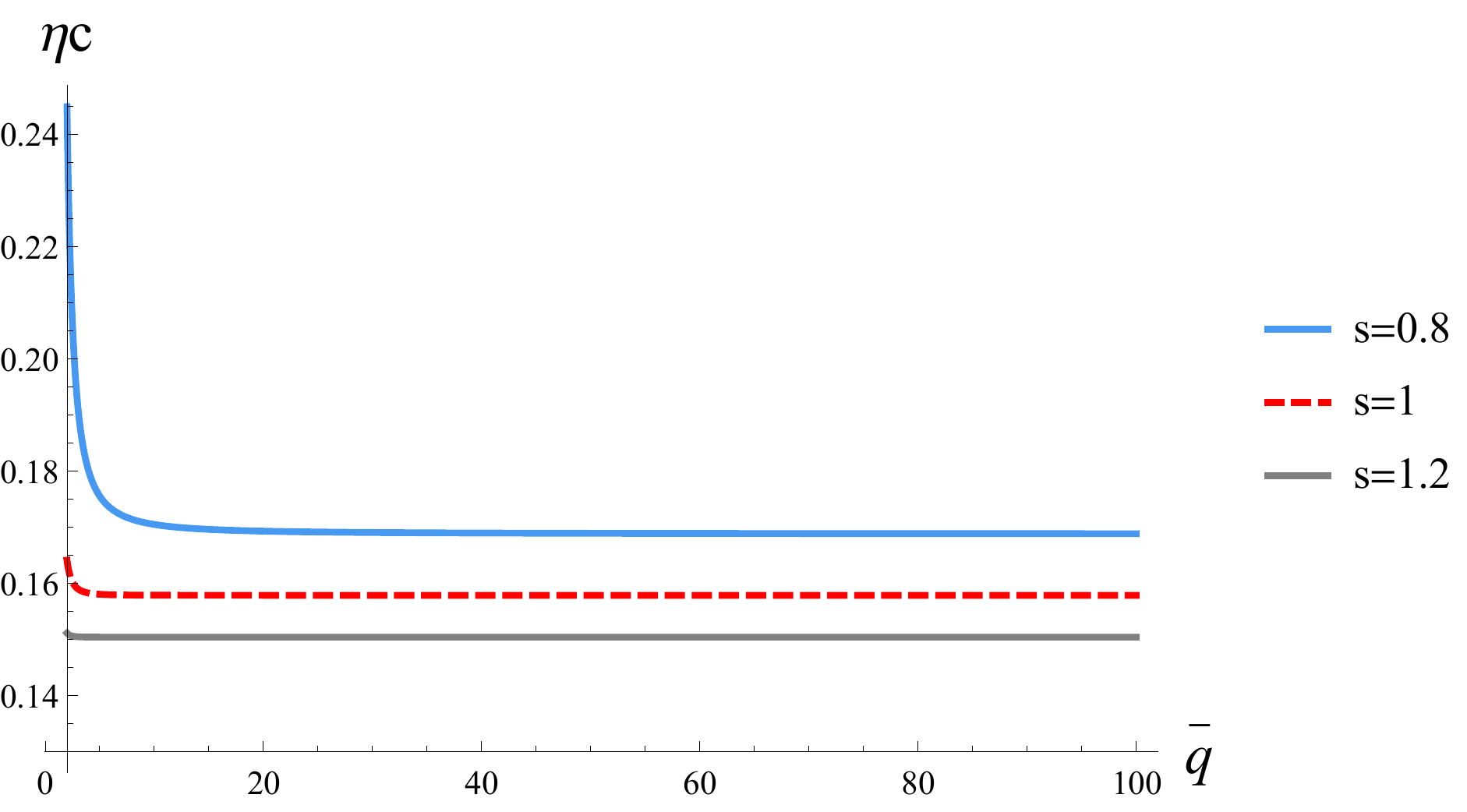} }
 			\caption{ The behaviour of (a) $\eta$ and  (b) $ \eta_{\rm C}$    with  charge $\bar{q}$ at different $s$.}   \label{fig:PMI plots of eta and etac}
 		}
 	\end{center}
 
 \end{figure}
 
  \begin{figure}[h]
    	\begin{center}
  		{\centering
  			\subfloat[]{\includegraphics[width=2.7in]{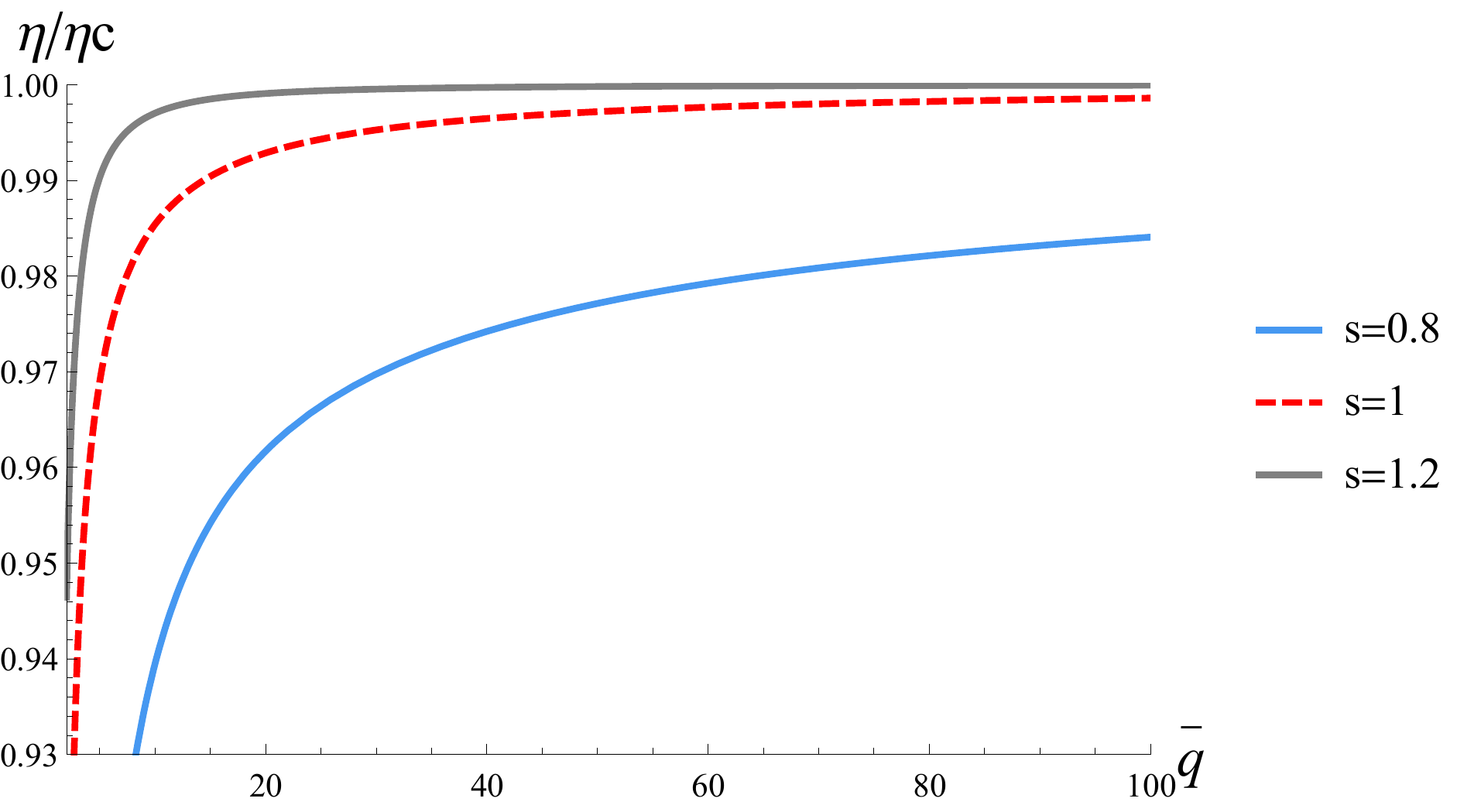} } \hspace {1.0cm}
  			\subfloat[]{\includegraphics[width=3in]{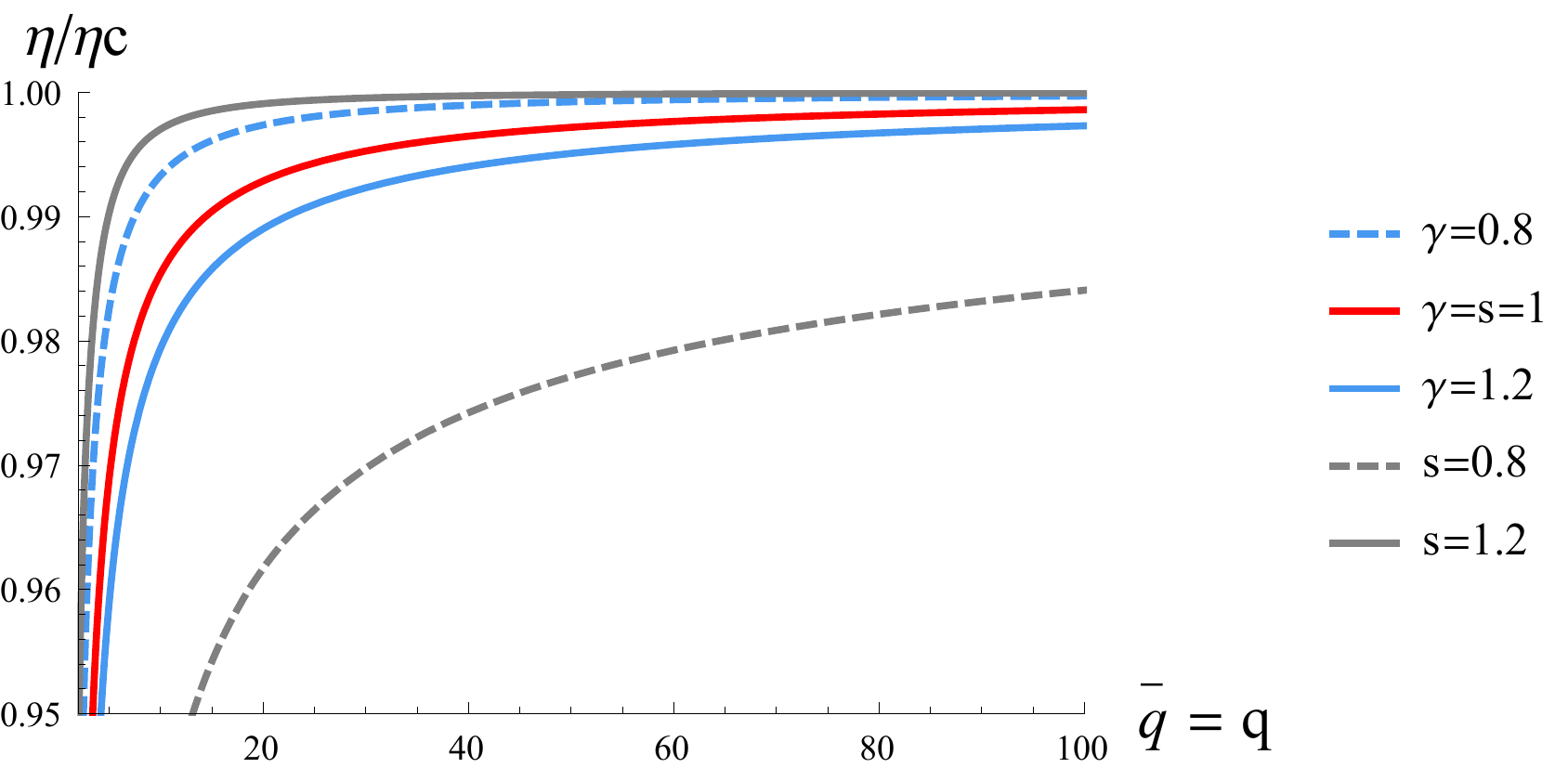} }
  			\caption{  (a) The behaviour of the ratio $\eta /\eta_{\rm C}$    with  charge $\bar{q}$ at different $s$, (b) A comparative plot of the ratio $\eta /\eta_{\rm C}$ between Yang-Mills and Maxwell charged black holes for different  $\gamma$ and $s$.}   \label{fig: comparative plot of YM and PMI}
  		}
  	\end{center}
    \end{figure}
 Figure (\ref{fig:PMI plots of eta and etac}) contains the variation of efficiency of the engine together with the Carnot efficiency as the non-linearity parameter $s$ is varied.
 Figure (\ref{fig: comparative plot of YM and PMI}) reveals that, in contrast to the Yang-Mills case, the approach of the ratio $\eta/\eta_{\rm C} $ towards unity is high for higher values of $s$.  On comparison of the ratio $\eta/\eta_{\rm C} $ between Yang-Mills charge and Maxwell charge, both the systems coincide when the power $\gamma =s=1$, where as Maxwell (Yang-Mills) dominates over Yang-Mills (Maxwell) when $\gamma=s> (<) 1$. 
 \vspace{0.5cm} \\
 \noindent 
These critical heat engines at large charge  are proposed as examples of low temperature heat engines (since $T_{\rm cr} \sim 1/q^{\kappa}$  (or $1/\bar{q}^b$)) similar to~\cite{Johnson:2017hxu,Bhamidipati:2017nau}.  
It is worth mentioning here that the series of $\eta$ (and $\eta_{\rm C} $) at large $q$ and $\bar{q}$ converge to the values $\Big(\frac{1-4\gamma}{1-20\gamma}\Big)$ and $\Big(\frac{1+2s}{1+18s}\Big)$, respectively. Interestingly, these convergent points satisfy the relation given in~\cite{Bhamidipati:2017nau}, summarized as,
\begin{equation}
\text{Convergent point of} \, \, \eta \, \, \text{and} \, \, \eta_{\rm C} = \frac{\rho_{cr}}{2 + \rho_{cr}} \, ,
\end{equation} 
where $\rho_{cr}$ is the critical ratio~\cite{Zhang:2014eap,Hendi:2012um}:
\begin{equation}
\rho_{cr} = \Bigg\{
\begin{array}{ccc}
\frac{4\gamma-1}{8\gamma}, & \text{YM charged black hole} \\ \\
\frac{2s+1}{8s}, & \text{Maxwell charged black hole} \ .	
\end{array}
\end{equation}
  
  %%%%%%%%%%%
\section{Motion of Test Particles in Critical Black Holes}
%%%%%%%%%%%%%%

%\section{Effective Potential}
\noindent
Let us now study the critical region by starting from the metric function of Power Yang-Mills black hole with critical values inserted. It is interesting to further explore the physcs of $q$ interacting
constituent degrees of freedom bound together by a potential. This can be studied by analyzing the motion of a
point particle of mass $\mu$ and charge $e$ moving in the
background of the critical charged black hole, in probe approximation, where the potential comes from the black hole geometry.  This follows from the idea of a  coupled interacting system approaching Carnot efficiency at criticality~\cite{Johnson:2017hxu,PhysRevLett.114.050601,power_of_a_critical_heat}. The corresponding study for power Maxwell black holes was recently presented in~\cite{Hendi:2018wib} and it would be nice to have a comparison of the two cases, following the comparison of effeciency of these systems in the last section. The critical metric function is:
\begin{equation}
Y(r)=1-\frac{2m_{cr}}{r}-\frac{r^2}{l_{cr}^2}+\frac{(2q^2)^\gamma}{2(4\gamma-3)r^{(4\gamma-2)}}\ ,
\label{eq:metric function cr}
\end{equation}
Other critical parameters such as mass and cosmological constant parameter can be found to be:
\begin{eqnarray}
&& m_{\rm cr} = \left\{\frac{1}{3} \left(3 \left(2^{\gamma } \gamma  (4 \gamma -1)\right)^{\frac{1}{4 \gamma -2}}
   q^{\frac{\gamma }{2 \gamma -1}}+\frac{(2 \gamma -1) \left(2^{\gamma } \gamma  (4 \gamma
   -1)\right)^{\frac{1}{4 \gamma -2}} q^{\frac{\gamma }{2 \gamma -1}}}{2 \gamma } \right.\right. \nonumber \\
 &&\qquad ~~~~~~ \left. \left. +\frac{3\ 2^{\gamma
   } \left(q^2\right)^{\gamma } \left(\left(2^{\gamma } \gamma  (4 \gamma -1)\right)^{\frac{1}{4
   \gamma -2}} q^{\frac{\gamma }{2 \gamma -1}}\right)^{3-4 \gamma }}{8 \gamma -6}\right)\right\}  \\
&&  l^2_{\rm cr} =  \left\{\frac{6 \gamma  \left(2^{\gamma } \gamma  (4 \gamma -1)\right)^{\frac{2}{4 \gamma -2}}
   q^{\frac{2 \gamma }{2 \gamma -1}}}{2 \gamma -1}\right\}
\end{eqnarray}
We note that the above parameters reduce to standard values corresponding to Reissner-Norsdr\''{o}m charged black holes for the case $\gamma=1$~\cite{Johnson:2017asf}.
%%%%%%%%%%%
 \begin{figure}[h]
 
 	\begin{center}
 		{\centering
 			\subfloat[]{\includegraphics[width=2.8in]{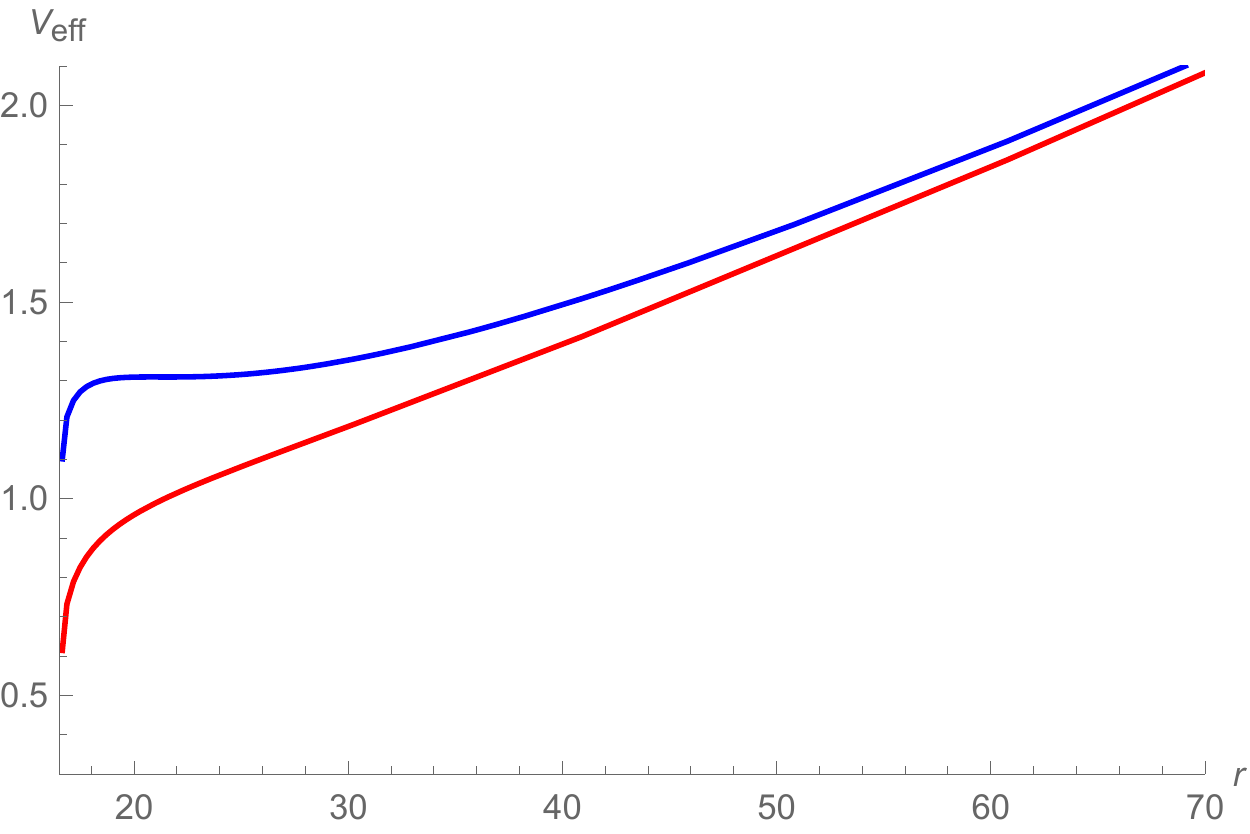} } \hspace {1.0cm}
 			\subfloat[]{\includegraphics[width=2.8in]{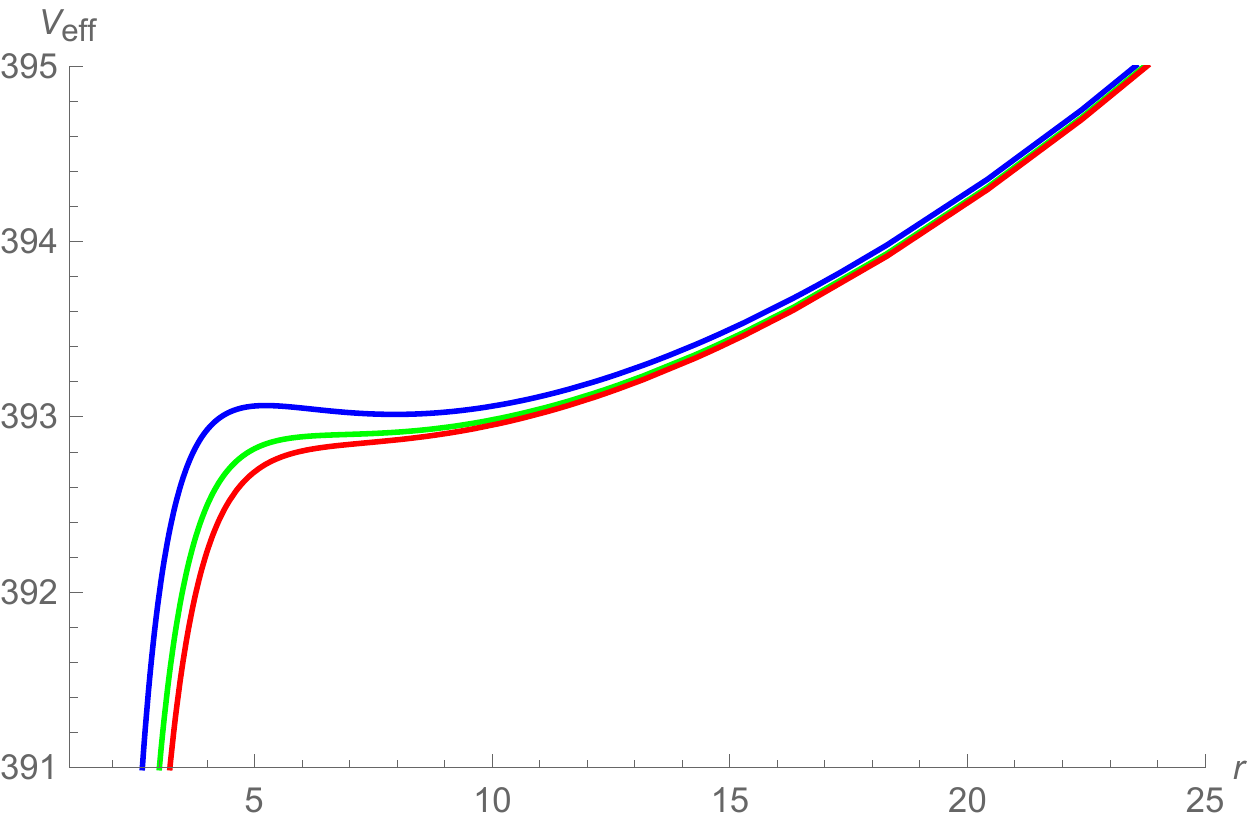} }
 			\caption{Effective potential plotted for (a) different mass to charge ratios, shows a minimum for $\mu/e=0.16$(Blue). The minimum disappears for $\mu/e=0.2$(Red). Values of other parameters are $\gamma=1.2, q=10, L=0$. (b) different values of non-linearity parameter $\gamma=0.9$(Blue), $\gamma=1$(Green), $\gamma=5$ (Red),  for $ q=1, \mu/e=4.$  }   \label{fig:Veff_r}
 		}
 		% M=2, L=200, \Lambda= - 11.5,
 	\end{center}
 \end{figure}

%%%%%%%%%%%
 \begin{figure}[h]
 
 	\begin{center}
 		{\centering
 			\subfloat[]{\includegraphics[width=2.8in]{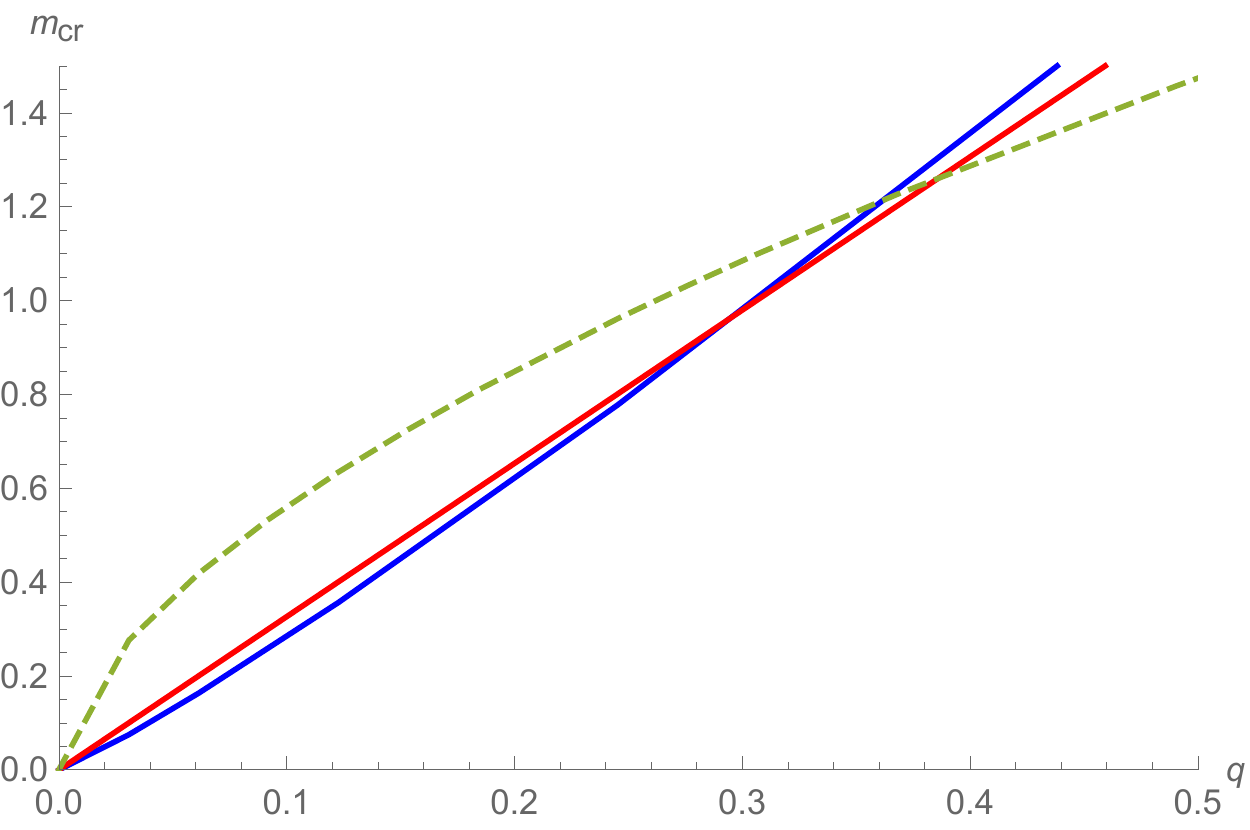} } \hspace {1.0cm}
 			\subfloat[]{\includegraphics[width=2.8in]{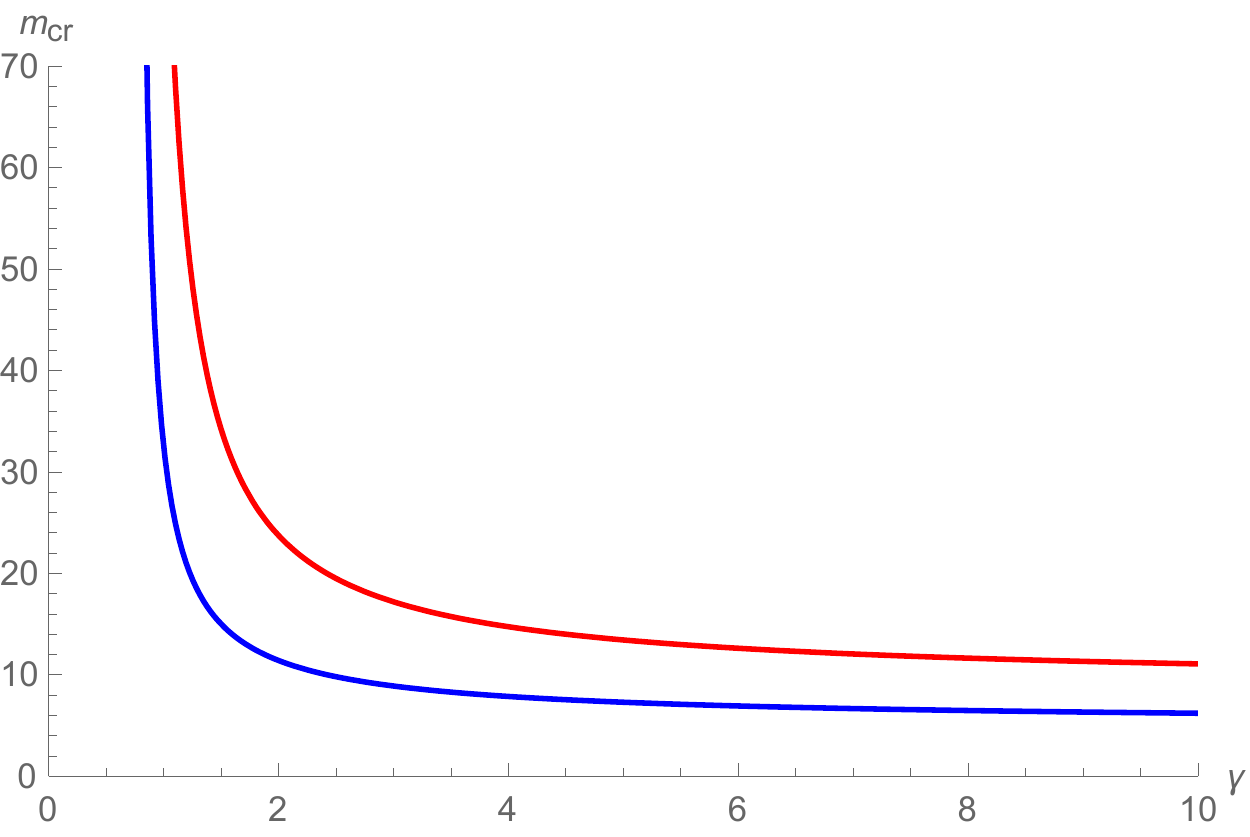} }
 			\caption{ Variation of critical mass with (a) charge $q$: for non-linearity parameter values $\gamma=0.9$(Blue), $\gamma= 1$(red), $\gamma=3$(Dashed) (b) with non-linearity parameter $\gamma$ for $q=10$ (blue), $q=30$(red).}   \label{fig:mcr_q}
 		}
 	\end{center}
 
 \end{figure}
 %%%%%%%%%%%%%%%%%%
 \begin{figure}[h!]
 	%\begin{wrapfigure}{r}{0.3\textwidth}
 	\begin{center}
 		{\centering
 		\subfloat[]	{\includegraphics[width=2.6in]{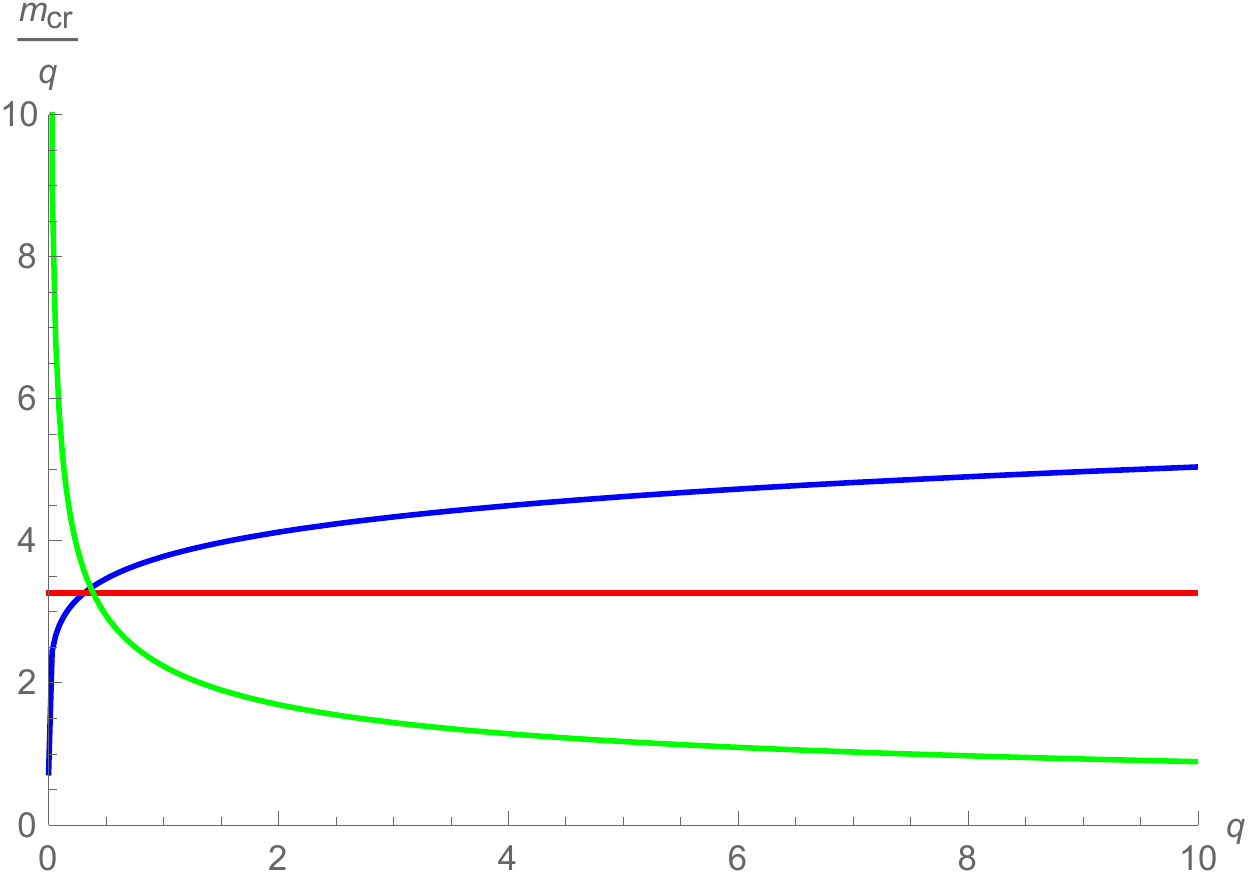} } \hspace {1.0cm} 		
 			\subfloat[]{\includegraphics[width=2.6in]{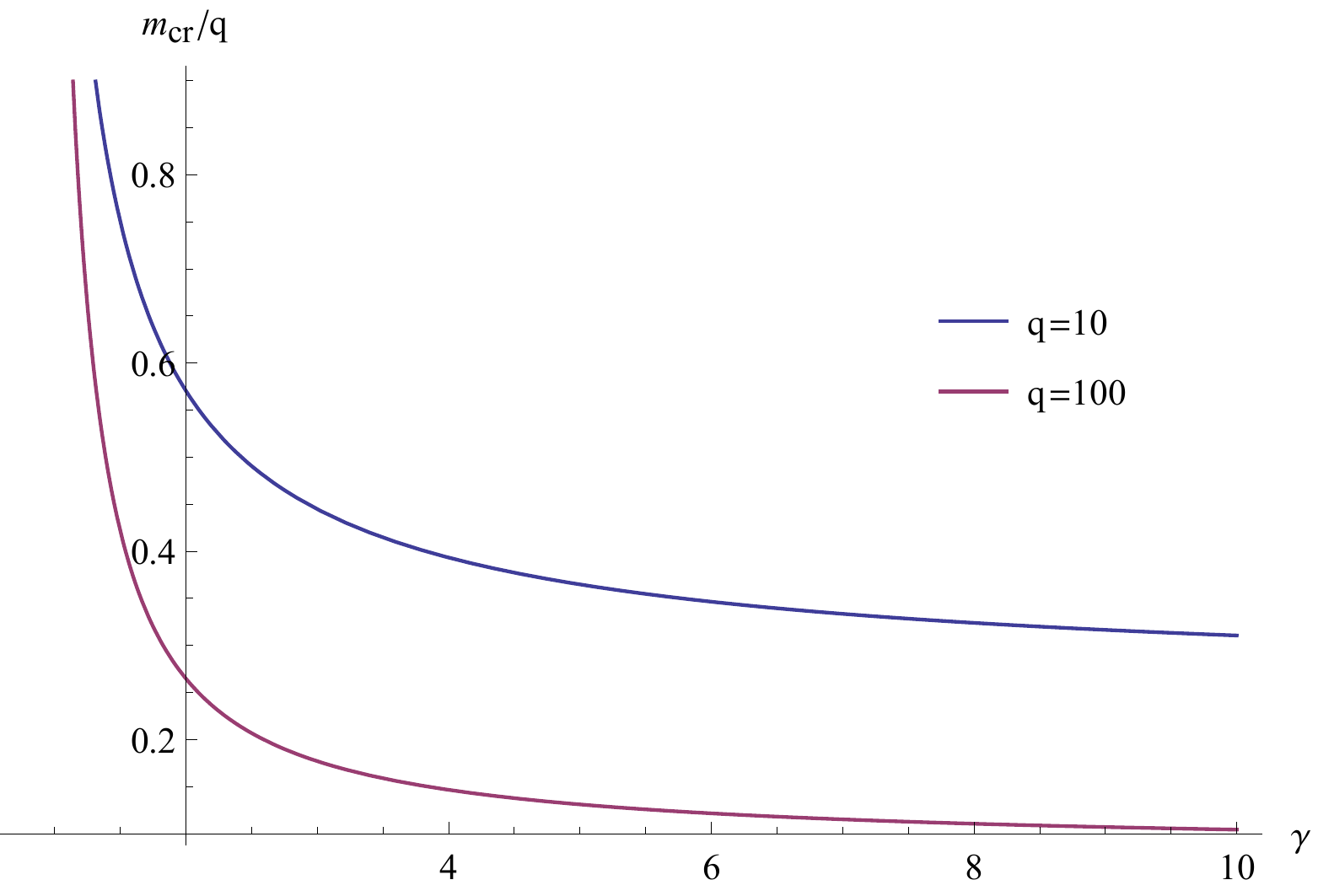} }
 			
 			\caption{Mass to charge ratio $m_{cr}/q$ plotted (a) with respect to charge $q$: for $\gamma=0.9$(Blue), $\gamma=1$(Red), $\gamma=3$(Green) (b) with respect to $\gamma$.  }   \label{fig:mcrbyq_q}
 		}
 	\end{center}
 	% \end{wrapfigure}
 \end{figure}
 %%%%%%%%%
 \noindent
Following the methods in~\cite{Chandrasekhar1984,doi:10.1142/S0217732311037261,Johnson:2017asf,Bhamidipati:2017nau}, the effective potential can be noted to be
\begin{equation}
V_{\rm eff}(r) = e\, \frac{2^{(\gamma-1)}q^{(2\gamma-1)}}{(4\gamma-3)}\gamma r^{(3-4\gamma)} +  \sqrt{Y_{\rm cr}(r)}\sqrt{\mu^2+\frac{L^2}{r^2}}\ ,
\end{equation}
where $L$ is the angular momentum of the particle. This is plotted in figure-(\ref{fig:Veff_r}) showing an attractive and binding behavior. The presence of a local minimum depends on fine tuned values of mass to charge ratio of the test particle. As seen from the figure-(\ref{fig:Veff_r}a) and also summarized in Table-1, this potential rises at large $r$ due to the presence of the 
cosmological constant term. Existence of a local minimum $r_{min} > r_{cr}$ depends upon the values of $\mu, e, L$ and the non-linearity parameter $ \gamma$.
  %%%%%%%%%%%%%%%

\begin{table}[tbp]
	{\bf{Table 1: To produce a attractive minimum in the effective potential: Minimum value of mass to charge ratio of particle required, for a given value of nonlinearity parameter $\gamma$}}\\
	\vskip 0.1cm
    \centering
    \begin{tabular}[t]{|c|c|c|}
 %       \toprule
        \hline\hline $\gamma$ & $\mu/e$ & $V_{\rm min}$ \\
        \hline\hline $ 0.9$ & $0.526 $ & $2.317 $  \\
               \hline\hline $0.95 $ & $0.37 $ & $ 1.964$  \\
                        \hline\hline $1$ & $0.25 $ & $1.855 $  \\
                                \hline\hline $1.05 $ & $0.2 $ & $1.629 $  \\
                                        \hline\hline $1.1 $ & $0.167 $ & $1.47 $  \\
                                         \hline\hline $1.15 $ & $0.134 $ & $1.389 $  \\
                                           \hline\hline $1.2 $ & $0.112 $ & $1.309 $  \\
                                             \hline\hline $1.25 $ & $ 0.093$ & $1.253 $  \\
                                               \hline\hline $1.3 $ & $ 0.076$ & $1.214 $  \\
         \hline
    \end{tabular}\hfill%
\end{table}

\noindent
It is interesting to study the behavior of critical mass, cosmological constant and critical radius as a function of the Power-Yang-Mills non-linearity parameter $\gamma$. Critical mass for instance is plotted in figure-(\ref{fig:mcr_q}) for different choices of $\gamma$ and charge $q$ (not shown are plots of other critical parameters, e.g., $l_{cr}$ and $r_{cr}$ etc., which are qualitatively similar.). The mass to charge ratio of black holes $m_{cr}/q$ shows an interesting behavior, plotted in figure-(\ref{fig:mcrbyq_q}). For $\gamma >1$,  $m_{cr}/q$ is a decreasing function of $q$, where as, it is an increasing function for values of $\gamma <1$. This behavior of  mass to charge ratio is opposite to the one noted in the case of Power-Maxwell's black holes in~\cite{Hendi:2018wib} and reminiscent of the contrast in behavior of efficiency seen in both the systems. Now, we study the near horizon geometry for power-Yang-Mills  critical black holes. By writing $r=r_++\epsilon\sigma$ and $t= \tau/\epsilon$ (where $\epsilon$ is small) in the neighbourhood of the horizon,
the metric (\ref{eq:staticform}) becomes(to first order in $\epsilon$)~\cite{Johnson:2017asf}:
\begin{equation}
ds^2 = -\Big(\frac{4\pi T}{\epsilon}\Big)\sigma d\tau^2+ \Big(\frac{\epsilon}{4\pi T}\Big)\frac{d\sigma^2}{\sigma } + (r_+^2 + 2\epsilon \sigma r_+)d\Omega_2^2.
\end{equation}
where, $Y(r)\vert_{\rm{r=(r_++\epsilon \sigma)}} = \epsilon \sigma Y'(r_+)$ and $Y'(r_+)=\frac{dY(r)}{dr}\vert_{\rm{r=r_+}} =4\pi T$ (see equation (\ref{equ:T})).
\\ Now, for the critical black hole, $ T = T_{cr}$ and $r_+ = r_{cr}$ then the metric becomes:
\begin{equation}
ds^2 = -\Big(\frac{4\pi T_{cr}}{\epsilon}\Big)\sigma d\tau^2+ \Big(\frac{\epsilon}{4\pi T_{cr}}\Big)\frac{d\sigma^2}{\sigma } + (r_{cr}^2 + 2\epsilon \sigma r_{cr})d\Omega_2^2.
\end{equation}
The near horizon limit is obtained by taking $\epsilon\to0$ while at the same time taking the large $q$ limit by holding  $\epsilon q^\kappa$ fixed, then
\begin{equation}
 ds^2 = -{(4\pi {\widetilde T}_{\rm cr})\,\,\sigma}d\tau^2+\frac{1}{(4\pi {\widetilde T}_{\rm cr})}\frac{d\sigma^2}{\sigma }+d{\mathbb R}^{2}.
\end{equation}
 Here, ${\widetilde T}_{\rm cr}$ = $T_{\rm cr}(\widetilde{q})$,  $\widetilde{q} = q\epsilon^{\frac{1}{\kappa}}$ and also,  $\Lambda=0$.
 In this limit of large $q$, where $r_{\rm cr}$ diverges (from eqn. \ref{eqn:YM critical point}), the metric on $S^2$ having the radius $r_{\rm cr}$ can be approximated with the flat metric $d{\mathbb R}^{2}=dx_1^2+dx_2^2\ $. Thus, this double limit generates a 4-dimensional Rindler space-time with zero cosmological constant, which is a fully decoupled space-time, which has been discussed in recent literature~\cite{Johnson:2017asf,Bhamidipati:2017nau,Hendi:2018wib}.\\

%%%%%%%%%%%%%
\section*{Appendix}
%%%%%%%%%%%%%
 When we set the Yang-Mills power $\gamma$ to be unity, the  physical quantities become:	
 \vspace{0.2cm}
 \\ The critical point turns out to be:
 \begin{eqnarray}
 T_{\rm cr}&=& \frac{1}{3\pi\sqrt{6} q} \ , \nonumber \\
 p_{\rm cr} &=& \frac{1}{96\pi  q^2} \ , \nonumber \\ 
 V_{\rm cr}&=& 8\sqrt{6}\pi q^3\ , \nonumber \\
 \text{and} \, \, \, \, r_{\rm cr}&=& \sqrt{6} q \ . 
 \end{eqnarray}
 Specific heat at constant  pressure  is given by:
 
 \begin{equation}
 C_p = \frac{2S(8pS^2+S-\pi q^2)}{(8pS^2-S+3\pi q^2)}\ .
 \end{equation}  
 
 The work done $W$ and the large $q$ expansions of  $Q_H$ and efficiency $\eta$ are:
 \begin{eqnarray}
 W &=& \frac{1}{4\sqrt{6}}, \nonumber \\
 Q_H &=&  \frac{19\sqrt{6}}{72} + \frac{\sqrt{6}}{27}\Big(\frac{1}{q}\Big) + \frac{4\sqrt{6}}{243}\Big(\frac{1}{q^2}\Big) + O \Big(\frac{1}{q^3}\Big), \nonumber \\
 \eta &=& \frac{3}{19} - \frac{8}{361}\Big(\frac{1}{q}\Big) - \frac{416}{61731}\Big(\frac{1}{q^2}\Big) - \frac{3286}{1172889}\Big(\frac{1}{q^3}\Big) + O \Big(\frac{1}{q^4}\Big),
 \end{eqnarray}

 \begin{figure}[h!]
 	%\begin{wrapfigure}{r}{0.3\textwidth}
 	\begin{center}
 		{\centering
 			{\includegraphics[width=2.6in]{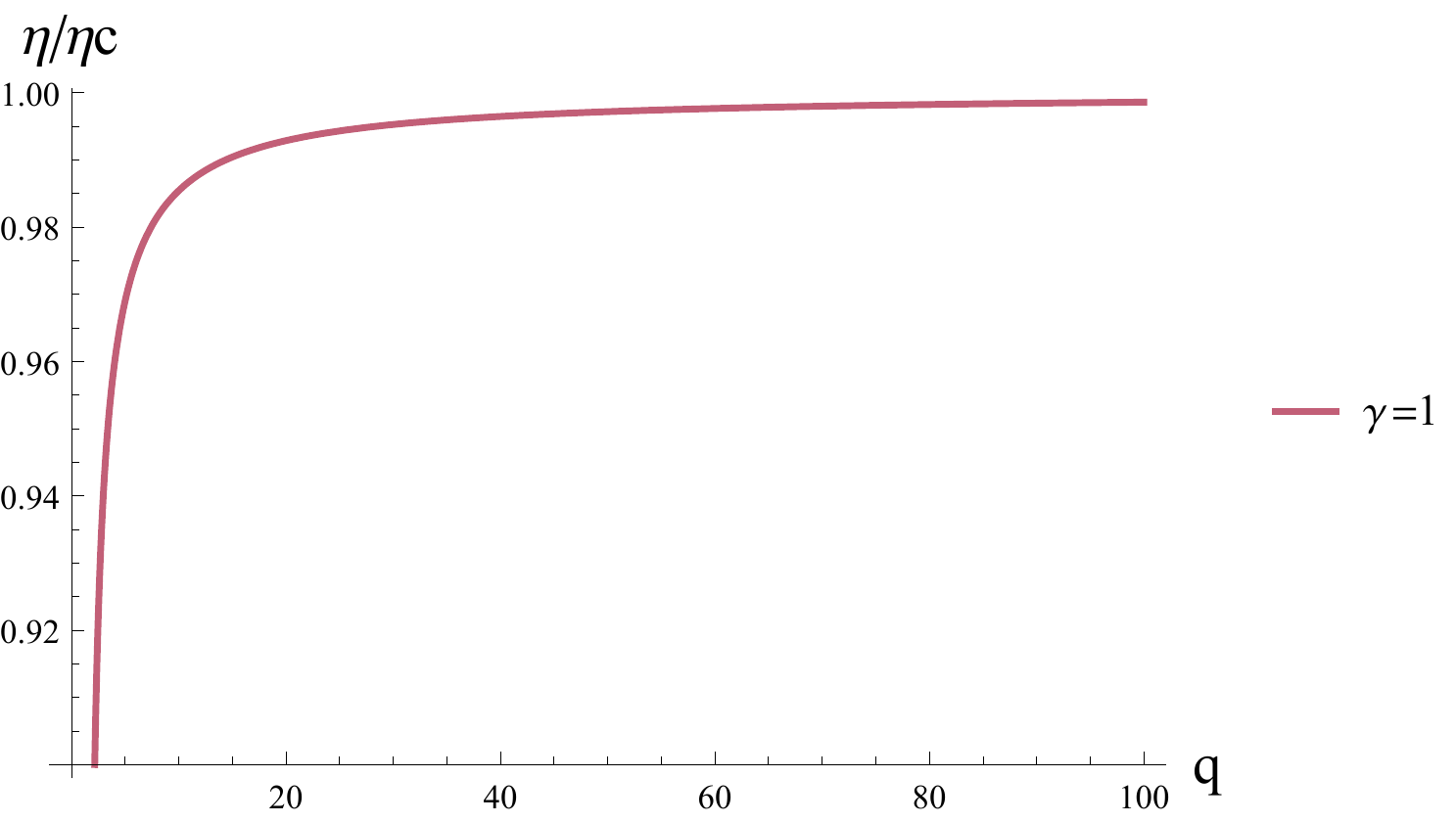} } 
 			
 			\caption{  The behaviour of the ratio $\eta/\eta_{\rm C}$ with $q$ at $\gamma=1.$  }   \label{fig: YM plot of ita by itac for gamma=1}
 		}
 	\end{center}
 	% \end{wrapfigure}
 \end{figure}
 
 while, the temperature $T_H$ and the large $q$ expansions of  $T_C$ and Carnot efficiency $\eta_{\rm C}^{\phantom{C}}$ are:
 \begin{eqnarray}
 T_H &=& \frac{19\sqrt{6}}{288\pi q}, \nonumber \\
 T_C &=&  \frac{\sqrt{6}}{18\pi q} - \frac{\sqrt{6}}{972\pi}\Big(\frac{1}{q^4}\Big) - \frac{7\sqrt{6}}{3888\pi}\Big(\frac{1}{q^5}\Big) + O \Big(\frac{1}{q^6}\Big), \nonumber \\
 \eta_{\rm C}^{\phantom{C}} &=& \frac{3}{19} + \frac{8}{513}\Big(\frac{1}{q^3}\Big) + \frac{14}{513}\Big(\frac{1}{q^4}\Big) + O \Big(\frac{1}{q^5}\Big).
 \end{eqnarray}
 Moreover, the time $\widehat{\tau} \sim q^2$ and,  
 \begin{equation} 
 \frac{\eta(\eta_{\rm C}-\eta)}{T_C}  =   \frac{72\pi\sqrt{6}}{6859} + \frac{224\pi\sqrt{6}}{130321}\Big(\frac{1}{q}\Big)  + O\Big( {\frac{1}{q^2}}\Big) \quad \, \, \text{(at  large $q$)}.
 \end{equation}

 Therefore, these results are exactly same as those of the critical heat engine for 4-dimensional Reissner-Nordstrom (RN)  black hole (with $L=1$, there)~\cite{Johnson:2017hxu}, where the Yang-Mills charge plays the role of the Maxwell charge.

%%%%%%%%%%%%%%%%%%%%%%%%%%%%%%%%%%%%%%%%%%%%%%%%%%%%%%%%%%%%%%%%%%%%%%%%%%%%%%%%% 

\bibliographystyle{apsrev4-1}

\bibliography{PYM}

\end{document}